\documentclass[useAMS,usenatbib]{mn2e}
\usepackage[english]{babel}
\usepackage{graphicx}
\usepackage{amssymb}

\title[Methods in chaos detection]{Chaos detection tools: application to a self--consistent triaxial model}

\author[N. P. Maffione et al.]{N. P. Maffione$^{1}$\thanks{E-mail:
nmaffione@fcaglp.unlp.edu.ar (NPM); ldarriba@fcaglp.unlp.edu.ar (LAD); pmc@fcaglp.unlp.edu.ar (PMC); giordano@fcaglp.unlp.edu.ar (CMG)}, L. A. Darriba$^{1}$\footnotemark[1], P. M. Cincotta$^{1}$\footnotemark[1] and C. M. Giordano$^{1}$\footnotemark[1]\\
$^{1}$Grupo de Caos en Sistemas Hamiltonianos, Facultad de Ciencias Astron\'omicas y Geof\'isicas, Universidad Nacional de La Plata and\\Instituto de Astrof\'isica de La Plata, CONICET-CCT La Plata, Paseo del Bosque s/n, La Plata, B1900FWA, Buenos Aires, Argentina}

\begin{document}

\date{Accepted. Received.}

\pagerange{\pageref{firstpage}--\pageref{lastpage}} \pubyear{--}

\maketitle

\label{firstpage}

\begin{abstract}
Together with the variational indicators of chaos, the spectral analysis methods have also achieved great popularity in the field of chaos detection. The former are based on the concept of local exponential divergence. The latter are based on the numerical analysis of some particular quantities of a single orbit, e.g. its frequency. In spite of having  totally different conceptual bases, they are  used for the very same goals such as, for instance, separating  the chaotic and the regular component. In fact, we show herein that the variational indicators serve to distinguish both components of a Hamiltonian system in a more reliable fashion than a spectral analysis method does.

We study two start spaces for different energy levels of a self--consistent triaxial stellar dynamical model by means of some selected variational indicators and a spectral analysis method. In order to select the appropriate tools for this paper, we extend previous studies where we make a comparison of several variational indicators on different scenarios. Herein, we compare the Average Power Law Exponent (APLE) and an alternative quantity given by the Mean Exponential Growth factor of Neary Orbits (MEGNO): the MEGNO's Slope Estimation of the largest Lyapunov Characteristic Exponent (SE{\it l}LCE). The spectral analysis method selected for the investigation is the Frequency Modified Fourier Transform (FMFT). 

Besides a comparative study of the APLE, the Fast Lyapunov Indicator (FLI), the Orthogonal Fast Lyapunov Indicator (OFLI) and the MEGNO/SE{\it l}LCE, we show that the SE{\it l}LCE could be an appropriate alternative to the MEGNO when studying large samples of initial conditions. The SE{\it l}LCE separates the chaotic and the regular components reliably and identifies the different levels of chaoticity. We show that the FMFT is not as reliable as the SE{\it l}LCE to describe clearly the chaotic domains in the experiments. We use the latter indicator as the main variational indicator to analyse the phase space portraits of the model under study. 
\end{abstract}


\section{Introduction}\label{intro}
The understanding of a realistic dynamical model is strongly related to the capability of identifying the chaotic and the regular nature of its orbits. 

One the one hand, one of the main features of chaotic orbits is their high sensitivity to the initial conditions (hereinafter i.c.), and the concept of local exponential divergence applies perfectly to identify such correspondence. On the other hand, the regular orbits do not show this behaviour. A seminal contribution to the field of chaos detection has been made by Lyapunov when he introduced the Lyapunov Characteristic Exponents, hereinafter LCEs \citep{O68,L92}. The LCEs are theoretical quantities that measure the local rate of exponential divergence. Therefore, the LCEs are appropriate to distinguish between chaotic and regular motion. 

The LCEs constitute the backbone of variational indicators of chaos (VICs). Moreover, their numerical implementation \citep{BGGS80,F84,TSR01,S10}, i.e. the Lyapunov Indicators (LIs), was a major contribution to the development of both fast and easy--to--compute new VICs. Nowadays, there is a plethora of VICs in the literature, such as the Smaller Alignment Index, SALI \citep{S01,SABV04,SESS04,BS06,C08,AVB10}, and its generalized version, the Generalized Alignment Index, GALI \citep{SBA07,SBA08,MA11}; the Mean Exponential Growth factor of Nearby Orbits, MEGNO \citep{CS00,CGS03,GC04,GKW05,GB08,LDV09,HCAG10,CLD11,MGC11}; the Fast Lyapunov Indicator, FLI \citep{FGL97,FLG97,FL98,FL00,LF01,GLF02,FL06,PFL08,TLF08,LGF10}; its variant, the Orthogonal Fast Lyapunov Indicator, OFLI \citep{FLFF02}; and a second order variant of the OFLI, the OFLI$_{TT}^2$ \citep{B05,BBS09,BBB09,BBS10}. Furthermore, we can include the VICs based on the properties of dynamical spectra like the invariant spectra of Stretching Numbers or Local Lyapunov Characteristic Numbers, LLCNs \citep{CVEFGLDL97,LF98}, the invariant spectra of Helicity or Twist Angles \citep{VC94,CV96,CV97,CVEFGLDL97,FL98,LF98,VCE98} and the Spectral Distance \citep{VCE99}. Finally, we include the Relative Lyapunov Indicator, RLI \citep{SEE00,SESF04,SSEPD07,SESS04}, which is not based on the evolution of the solution of the first variational equations as the others, but on the evolution of two different but very close orbits. 

Notwithstanding the large number of VICs cited, we could also include alternative methods, for instance, the Average Power Law Exponent, APLE \citep{LVE08}, which is based on the concept of Tsallis entropy; evolutionary algorithms \citep{PABV09}; and the record could indeed continue.  

Together with the VICs, the spectral analysis methods (SAMs), based on the numerical analysis of some particular quantities of a single orbit, e.g. its frequency, have also achieved great popularity in the field of chaos detection. Among the SAMs we find the method outlined by \citet{L90}: the Frequency Map Analysis, FMA \citep{LFC92,L93,PL96,PL98} and a modification of the latter in order to improve the precision in the computed frequencies and amplitudes, the Frequency Modified Fourier Transform, FMFT \citep{SN97} or other alternatives using Fast Fourier Transform (FFT) techniques \citep{MF95,FMBC05,MLCF10}, among others. As an easy identification of the important resonances (regions of phase space occupied by orbits with commensurable orbital frequencies) is possible with the SAMs, they are able to identify the major resonant orbit families. The relative importance of each family regarding the phase space can be assessed from the number of orbits associated with the particular resonance the family belongs to. That is, the SAMs are not only suitable to determine the resonance web like the VICs \citep{KK94,CGS03,FLG06,LVE08} but also to identify the individual resonances by their frequency vectors.

Though the conceptual bases of the VICs and the SAMs are totally different, many researchers use both types of techniques for similar goals. Nevertheless, we show herein that this seems to be a no reliable approach to study some important aspects of the dynamics of a Hamiltonian system. Moreover, in the literature there are plenty of evidence that the complementary use of both types of techniques results an efficient and solid way to gather dynamical information from a Hamiltonian system \citep{FLG97,FL98,GLF02,KV05,BBB09}. For instance, in \citet{FLG97} the authors compare the sensitivity of the FLI and the FMA on the standard map and on the \citet{HH64} potential. In \citet{GLF02} the authors use the FLI and the Analytically Filtered Fourier Analysis, AFFA \citep{GB01}, another example of SAM, to determine the value of the critical parameter at which the transition from the unfortunately called Nekhoroshev to the Chirikov regime occurs in a quasi--integrable Hamiltonian model and a standard $4$--dimensional map. In \citet{KV05}, the authors use a measure related with the LCEs, the alignment indices \citep{VCE98,VCE99,S01} and an index computed with the FMFT to study two different N--body models simulating elliptical galaxies.

One of the aspects that makes a research accurate is the reliability of the tools available for the analysis. This reliability is a measure of, for instance, the confidence and efficiency of such tools. Regarding chaos detection, we believe that it is on the diversity of the employed techniques that the fruitfulness of the research relies. Therefore, we started a series composed of four papers in order to achieve the best combination of tools to study a given dynamical problem. In a first paper, \citet{MGC11}, we test the MEGNO against the most widely used chaos detection technique in the literature, which is the LI, on a somewhat realistic problem: a self--consistent triaxial stellar (ScTS) dynamical model \citep{MCW05,CGM08}. In the second and third papers, \citet{MDCG11} (hereinafter M11) and \citet{DMCG12} (hereinafter D12), we make a comparison of most of the VICs mentioned earlier on rather simple problems, i.e. on mappings and on the model of \citet{HH64}, respectively. Our purpose here is to extend the previous comparisons of VICs to alternative techniques, including a spectral analysis method. Further, this last paper of the series is regarded as a closing report on the subject and the conclusions comprise the whole investigation. We continue with the ScTS model and study two start spaces for four different energy levels by means of several chaos detection tools. In order to select appropriate VICs for the study, we consider the results of M11 and D12 and add a rather new technique, the APLE, and an alternative quantity given by the MEGNO: the MEGNO's Slope Estimation of the largest LCE (SE{\it l}LCE) to their comparison. The MEGNO is part of the suggested VICs to compose the CIsF\footnote{CIsF: ``Chaos Indicators Function''. It is a minimal and efficient package of variational indicators to study a general Hamiltonian.} in D12. In fact, we are interested in studying if its variant, i.e. the SE{\it l}LCE, is more reliable than the MEGNO on its own.  

The study is organized as follows: in Section \ref{themodel} we describe the ScTS model and introduce the $p_{x_0}-p_{z_0}$ and the $x_0-z_0$ start spaces. In Section \ref{CIs&FMFT}, we briefly introduce the chaos detection techniques to be used within the reported research. We start with the VICs: the LI, the MEGNO and the SE{\it l}LCE (Section \ref{CIs&FMFT-S1}); the FLI and the OFLI (Section \ref{CIs&FMFT-S2}) and the APLE (Section \ref{CIs&FMFT-S3}). We finish with a short FMFT introduction (Section \ref{CIs&FMFT-S4}). In Section \ref{theFMFT}, we deal with the capability of the FMFT to resolve the different structures in a divided phase space (Section \ref{FMFT-CIS1}) and compare its performance with the SE{\it l}LCE in both start spaces of the ScTS model (Section \ref{FMFT-CIS2}). In Section \ref{STATIONARY}, we study the efficiency of several VICs on the $p_{x_0}-p_{z_0}$ start space to decide which ones are appropriate to investigate the dynamics of the ScTS model (Sections \ref{STATIONARYS1} and \ref{STATIONARYS2}). In Section \ref{thespaces} we analyse for four different energy surfaces, the $p_{x_0}-p_{z_0}$ (Section \ref{STATIONARYS4}) and the $x_0-z_0$ (Section \ref{STARTX0Z0}) start spaces by means of a complementary use of the VICs selected in the previous section. Finally, Section \ref{conclusions} summarizes the main conclusions of the whole investigation. 

As the extensive used of acronyms in the text might be overwhelming for the reader, in Table \ref{acronyms} we list in alphabetical order the most important acronyms used in this paper. 

\begin{table}
\centering
\caption{List of the most important acronyms in alphabetical order.}
\begin{tabular}{@{}cc@{}}
\hline
 Acronym & Definition \\
\hline 
 APLE & Average Power Law Exponent \\
 CIsF & Chaos Indicators Function \\
 FLI & Fast Lyapunov Indicator \\
 FMA & Frequency Map Analysis \\
 FMFT & Frequency Modified Fourier Transform \\
 LCEs & Lyapunov Characteristic Exponents \\
 LIs & Lyapunov Indicators \\
 MEGNO & Mean Exponential Growth factor of Nearby Orbits \\
 OFLI & Orthogonal Fast Lyapunov Indicator \\
 SE{\it l}LCE& Slope Estimation of the largest LCE \\
 ScTS model & Self--consistent Triaxial Stellar model \\
 VICs & Variational Indicators of Chaos \\
\hline
\end{tabular}
\label{acronyms}
\end{table}

\section{The model and the start spaces}\label{themodel}
In order to obtain a fairly realistic dynamical model of an elliptical galaxy, in \citet{MCW05} the authors follow the cold disipationaless collapse of 100000 particles randomly distributed following a $1/r$ law within a sphere. The velocities were randomly chosen from a spherical Gaussian distribution but only their tangential component was retained. After the system had relaxed, there remained 86818 particles resembling an elliptical galaxy: the ScTS model. The system observes the de Vaucouleurs law shown by \citet{MCW05} in their Fig. 2. The model reproduces many dynamical characteristics of real elliptical galaxies, such as mass distribution, flattening, triaxiality and rotation \citep{M06}. The ScTS model has a strong triaxiality and a flattening that increases from the border of the system to its center (see Table I in the same paper). The resulting triaxial potential has semi--axis $X,Y,Z$ satisfying the condition $X>Y>Z$, and its minimum, which is close to $-7$, matches the origin. The potential is less flattened than the mass distribution, as expected. See Table I in \citet{MCW05}.

The ScTS model is an $N$--body potential which is neither smooth nor stationary. As the authors needed to compute the orbits and their corresponding LIs, they froze the potential and represented it with a quadrupolar approximation. The equation that reproduces the potential is:

\[
V(\mathbf{x})=-f_{0}(\mathbf{x})-f_{x}(\mathbf{x})\cdot(x^{2}-y^{2})-f_{z}(\mathbf{x})\cdot(z^{2}-y^{2})
\]
with $\mathbf{x}=(x,y,z)$ and where

\begin{equation}
f_{n}(\mathbf{x})=\frac{\alpha_{n}}{\left[p_{n}^{a_{n}}+\delta_{n}^{a_{n}}\right]^{\frac{ac_{n}}{a_{n}}}},
\label{model1}
\end{equation}
with $p^{2}_{n}$ the square of the softened radius given by $p_{n}^2=x^2+y^2+z^2+\epsilon^2$ when $n=0$, or $p_{n}^2=x^2+y^2+z^2+2\cdot\epsilon^2$ for $n=x,z$, and $\alpha_{n}$, $\delta_{n}$, $a_{n}$, $ac_{n}$ are constants. 

The selected value for the softening parameter is $\epsilon\simeq 0.01$ for any $n$. The adopted values for the constants $\alpha_{n}$, $\delta_{n}$, $a_{n}$ and $ac_{n}$ are given in Table \ref{modeltable}. 

\begin{table}
\centering
\caption{Adopted values for the coefficients of the functions $f_n$ given by Eq. (\ref{model1}).}
\begin{tabular}{@{}ccccc@{}}
\hline
  & $\alpha$ & $a$ & $\delta$ & $ac$\\
\hline 
$n=0$ & $0.92012657$ & $1.15$ & $0.1340$  & $1.03766579$\\
$n=x$ & $0.08526504$ & $0.97$ & $0.1283$ & $ 4.61571581$\\
$n=z$ & $-0.05871011$ & $1.05$ & $0.1239$ & $4.42030943$\\
\hline
\end{tabular}
\label{modeltable}
\end{table}

The stationary character of the parameters given in Table \ref{modeltable} were  tested by performing several fits at different times after virialization, resulting in a precision of $0.1\%$. For further details on the ScTS model refer to \citet{MCW05,CGM08}. 

We have already used the same model to test the MEGNO in a previous study \citep{MGC11}.

The spaces of i.c. used in this investigation are the $p_{x_0}-p_{z_0}$ and the $x_0-z_0$ start spaces for four energy surfaces: $-0.1,-0.3,-0.5$ and $-0.7$. The $p_{x_0}-p_{z_0}$ start space is a set of i.c. of the form $(p_{x_0},p_{z_0})$ with $x_0=y_0=z_0=0$ while $p_{y_0}$ is restrained by the energy condition \citep{PL98}. The $x_0-z_0$ start space is a set of i.c. of the form $(x_0,z_0)$ with $y_0=0,\,p_{x_0}=p_{z_0}=0$ and $p_{y_0}$ is restrained by the energy condition \citep{S93}. On the one hand, the combination of both start spaces is an adequate choice to sample many of the different orbital families within the triaxial model \citep{S93,PL98}. On the other hand, the stable boxlets (i.e. resonant box orbits) which are centrophobic are not well represented in our $p_{x_0}-p_{z_0}$ start space.  

Therefore, the $p_{x_0}-p_{z_0}$ and the $x_0-z_0$ start spaces of the ScTS model seem to provide fairly realistic scenarios for testing many characteristics of both the VICs of the package and the FMFT. 

\section{The techniques}\label{CIs&FMFT}
 The aim of this section is to briefly introduce the techniques we will use in the forthcoming study. 
 
\subsection{The LI, the MEGNO and the SE{\it l}LCE}\label{CIs&FMFT-S1}
Considering a continuous dynamical system defined on a differentiable manifold $\mathcal{S}$, where $\mathbf{\Phi}^t(\mathbf{x})=\mathbf{x}(t)$ characterises the state of the system at time $t$, being the state of the system at time $t=0$, $\mathbf{x}(0)=\mathbf{x}_0$. The state of the system after two consecutive time steps $t$ and $t'$ will be given by the composition law: $\mathbf{\Phi}^{t+t'}=\mathbf{\Phi}^t\circ\mathbf{\Phi}^{t'}$.

Moreover, the tangent space of $\mathbf{x}$ maps onto the tangent space of $\mathbf{\Phi}^t(\mathbf{x})$ according to the operator $d_{\mathbf{x}}\mathbf{\Phi}^t$ and following the rule $\mathbf{w}(t)=d_{\mathbf{x}}\mathbf{\Phi}^t(\mathbf{w}(0))$, where $\mathbf{w}(0)$ is an initial deviation vector (hereinafter i.d.v.). The action of such operator at consecutive time intervals satisfies the equation: 

\[
d_{\mathbf{x}}\mathbf{\Phi}^{t+t'}=d_{\mathbf{\Phi}^{t'}(\mathbf{x})}\mathbf{\Phi}^t\circ d_{\mathbf{x}}\mathbf{\Phi}^{t'}.
\]

If we suppose that our manifold $\mathcal{S}$ has some norm denoted by $\|\cdot\|$, we can define the useful quantity:

\[\lambda_t(\mathbf{x})=\frac{\|d_{\mathbf{x}}\mathbf{\Phi}^t(\mathbf{w})\|}{\|\mathbf{w}\|}\]
called ``growth factor'' in the direction of $\mathbf{w}$. 

Let $\mathcal{H}({\mathbf{p}},{\mathbf{q}})$ with ${\mathbf{p}},\,{\mathbf{q}}\in \mathbb{R}^N$ be an $N$--dimensional Hamiltonian, that we suppose autonomous just for the sake of simplicity. Introducing the following notation:

\[{\mathbf{x}}=({\mathbf{p}},{\mathbf{q}})\in\mathbb{R}^{2N},\, \mathbf{f}(\mathbf{x})=(-\partial\mathcal{H}/\partial{\mathbf{q}},\ \partial\mathcal{H}/\partial{\mathbf{p}})\in\mathbb{R}^{2N},\]
the equations of motion can be written in a simple way as

\begin{equation}
\dot{\mathbf{x}}={\mathbf{f}}({\mathbf{x}}).
\label{megno1}
\end{equation}

Let $\gamma(\mathbf{x}_0;t)$ be an arc of an orbit of the flow (\ref{megno1}) over a compact energy surface: $M_h\subset\mathbb{R}^{2N}$, $M_h=\{{\mathbf{x}}: \mathcal{H}({\mathbf{p}},{\mathbf{q}})=h\}$, then

\[\gamma(\mathbf{x}_0;t)=\{{\mathbf{x}}({\mathbf{x}}_0;t'):{\mathbf{x}}_0\in M_h,\ 0\le t'< t\}.\]

We can gain fundamental information about the Hamiltonian flow in the neighborhood of any orbit $\gamma$ through the largest LCE: lLCE, defined as:

\begin{equation}
\chi_{\gamma}=\chi[\gamma(\mathbf{x}_0;t)]=\lim_{t\to\infty}\frac{1}{t}\ln\lambda_t[\gamma(\mathbf{x}_0;t)]
\label{megno2}
\end{equation}
with 

\[\lambda_t[\gamma(\mathbf{x}_0;t)]=\frac{\|d_{\gamma}\mathbf{\Phi}^t(\mathbf{w})\|}{\|\mathbf{w}\|}\] 
where $\|d_{\gamma}\mathbf{\Phi}^t(\mathbf{w})\|$ is an ``infinitesimal displacement'' from $\gamma$ at time $t$. The fact that the lLCE (and its truncated value, the so--called LI$=\lim_{t\to T}\frac{1}{t}\ln\lambda_t[\gamma(\mathbf{x}_0;t)]$ for T finite) measures the local mean exponential rate of divergence of nearby orbits is clearly understood when Eq. (\ref{megno2}) is written in an integral fashion:

\begin{equation}
\chi_{\gamma}=\lim_{t\to\infty}{1\over t}\int_0^t{\|\dot{d}_{\gamma}\mathbf{\Phi}^{t'}(\mathbf{w})\|
\over\|d_{\gamma}\mathbf{\Phi}^{t'}(\mathbf{w})\|}{\rm d}t'= \overline{\left(\|\dot{d}_{\gamma}\mathbf{\Phi}^{t'}(\mathbf{w})\|\over\|d_{\gamma}\mathbf{\Phi}^t(\mathbf{w})\|\right)},
\label{megno3}
\end{equation}
where the bar denotes time average. 

Finally, the orbit is classified as regular or chaotic if the LI tends to zero (as $\ln(t)/t$) or to a positive value, respectively (a detailed discussion on the theory and computation of the LCEs can be found in the review of \citet{S10}).

Now, we can introduce the quantity $Y[\gamma(\mathbf{x}_0;t)]$ through the expression:

\[Y_{\gamma}=Y[\gamma(\mathbf{x}_0;t)]={2\over t}\int_0^t{\|\dot{d}_{\gamma}\mathbf{\Phi}^{t'}(\mathbf{w})\|\over\|d_{\gamma}\mathbf{\Phi}^{t'}(\mathbf{w})\|}t'{\rm d}t',\]
which is related to the integral in Eq. (\ref{megno3}); i.e., in case of an exponential increase of $\|d_{\gamma}\mathbf{\Phi}^{t}(\mathbf{w})\|$, $\|d_{\gamma}\mathbf{\Phi}^{t}(\mathbf{w})\|=\|\mathbf{w}\|\cdot\exp(\chi_{\gamma} t)$, the quantity $Y_{\gamma}$ can be considered as a weighted variant of the integral in Eq. (\ref{megno3}). Notice that the quantity $\hat{\chi}_\gamma=Y_\gamma/t$ verifies $\hat{\chi}_\gamma\sim 2/t$ for regular motion and $\hat{\chi}_\gamma\sim\chi_\gamma$ for chaotic motion, with $t\to\infty$, which show that, in case of regular motion $\hat{\chi}_\gamma$ converges to 0 faster than $\chi_\gamma$ does (which goes to zero as $\ln t/t)$, while for chaotic motion both magnitudes approach the positive lLCE at a rather similar rate. 

Instead of using the instantaneous rate of increase, $\chi_{\gamma}$, we average the logarithm of the growth factor, $\ln(\lambda_t)=\chi_{\gamma} t$. Introducing the time average (MEGNO):

\[\overline{Y}_{\gamma}=\overline{Y}[\gamma(\mathbf{x}_0;t)]\equiv{1\over t}\int_0^tY_{\gamma}{\rm d}t',\]
we notice that the time evolution of the MEGNO can be briefly described in a suitable and unique expression for all kinds of motion. Indeed, its asymptotic behaviour can be summarized as follows:  

\begin{equation}
\overline{Y}_{\gamma}\approx a_{\gamma}t+b_{\gamma},
\label{megno4}
\end{equation} 
where $a_{\gamma}=0$ and $b_{\gamma}\approx 2$ for quasi--periodic motion, while $a_{\gamma}=\chi_{\gamma}/2$ and $b_{\gamma}\approx 0$ for irregular, stochastic motion. Deviations from the value $b_{\gamma}\approx 2$ (and $a_{\gamma}=0$) indicate that $\gamma$ is close to some particular objects in phase space, being $b_{\gamma}\lesssim 2$ or $b_{\gamma}\gtrsim 2$ for stable periodic orbits (resonant elliptic tori), or unstable periodic orbits (hyperbolic tori), respectively.

Refer to Fig. 1 (c) and (d) from \citet{CGS03} to see the general behaviour of the MEGNO for regular and chaotic orbits.

It is possible to estimate the LI of the orbit from Eq. (\ref{megno4}) if we apply a simple linear least--squares fitting on the MEGNO \citep{CS00,CGS03}. This estimation is the earlier mentioned SE{\it l}LCE indicator. The least--squares fitting at the end of the integration process uses the last 80\% of the points in order to avoid the initial transient. Thus, the SE{\it l}LCE almost considers the full history of the orbit which makes it a very sensitive VIC. 

The performances of the MEGNO's time evolution curves to characterise the different levels of stability and chaoticity of the orbits shown in M11 and D12 was excellent. On the other hand, the description of the regular components in a divided phase space by means of the final values of the MEGNO (i.e. the values of the indicator at the end of the total integration time), was not satisfactory. The SE{\it l}LCE seems to be an appropriate alternative to solve the problem.

\begin{figure}
\begin{tabular}{c}
\resizebox{75mm}{!}{\includegraphics{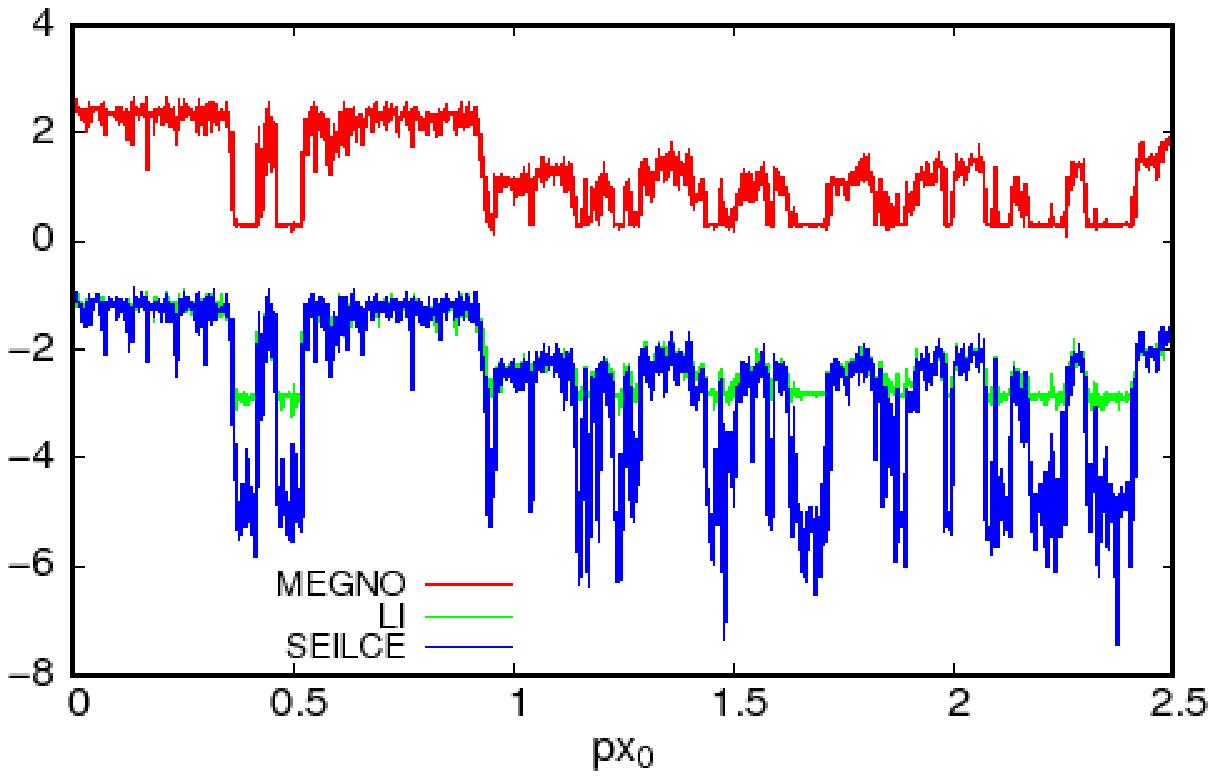}}\\
\resizebox{75mm}{!}{\includegraphics{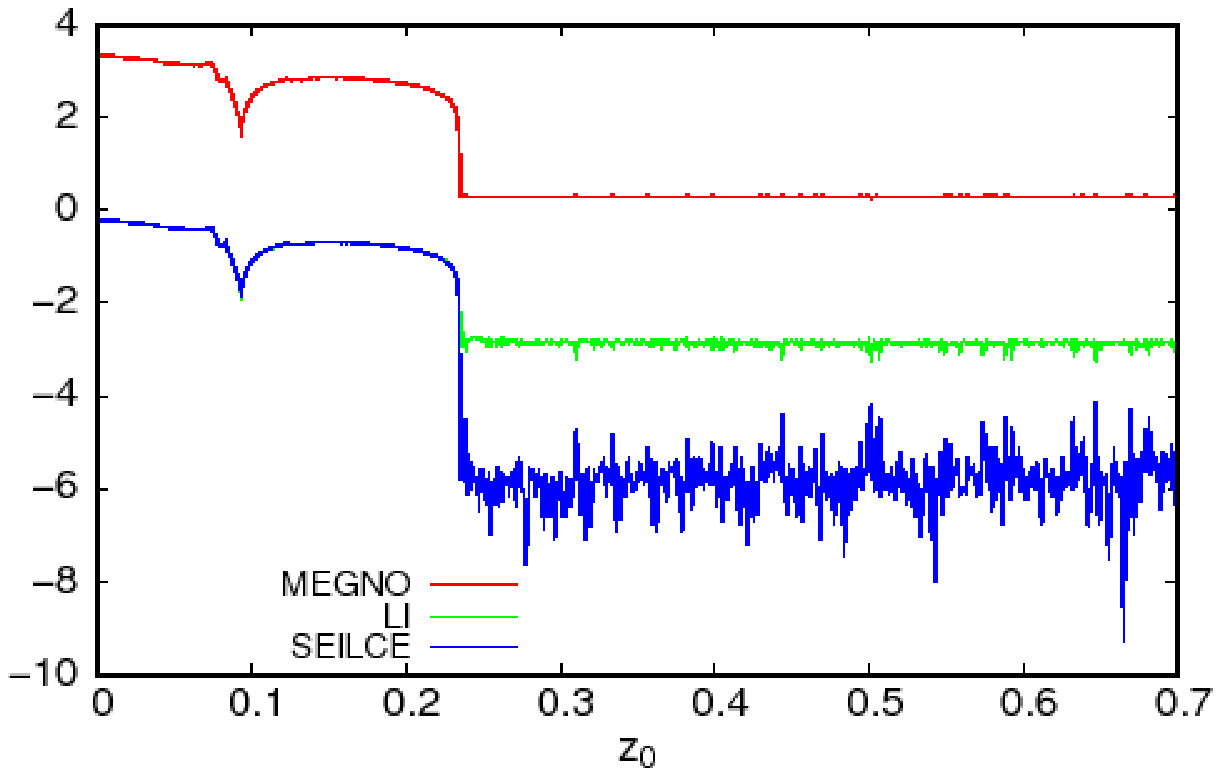}}
\end{tabular}
\caption{Final values of the LI, the MEGNO and the SE{\it l}LCE. The i.c. were taken in the $p_{x_0}-p_{z_0}$ start space (top panel) and in the $x_0-z_0$ start space (bottom panel). Both sets of i.c. were taken on the energy surface $-0.7$ and the VICs were integrated for $7\times 10^3$ u.t. The values of the indicators are in logarithmic scale.}
\label{CIs&FMFT-2}
\end{figure}

Fig. \ref{CIs&FMFT-2} shows the performances of the LI, the MEGNO and the SE{\it l}LCE for two sets of i.c., one of them located in the $p_{x_0}-p_{z_0}$ start space (top panel) and the other set located in the $x_0-z_0$ start space (bottom panel). In order to ensure that the VICs are well computed for a given orbit, the total integration time should verify $T>>T_c(E)$, where the function $T_c(E)$ is some characteristic time--scale which depends on the energy surface $E$. The function $T_c(E)$ is taken as the period of the stable X--axis periodic orbit and we consider the energy surface $E=-0.7$ for both sets of i.c. ($T_c(-0.7)\sim 7$ units of time --hereinafter u.t.-- see \citet{MGC11}). Therefrom, on fixing the condition $T\geq10^3\,T_c(E)$ to obtain reliable values for the VICs, we conclude that a total integration time of $7\times10^3$ u.t. would be appropriate for the experiment \citep{MGC11}. In the $p_{x_0}-p_{z_0}$ start space, the set is composed of 790 i.c. of the form $p_{x_0}\in[0:2.5]$ and $p_{z_0}=0$. In the $x_0-z_0$ start space, the set is composed of 646 i.c. of the form $z_{0}\in[0:0.7]$ and $x_{0}=0$. 

In Fig. \ref{CIs&FMFT-2}, the SE{\it l}LCE and the LI show very similar final values for chaotic orbits. The MEGNO grows exponentially fast and reaches higher final values ($\gtrsim 2$) than the SE{\it l}LCE or the LI. The MEGNO clearly distinguishes between different levels of chaoticity. Consequently, the SE{\it l}LCE seems to be an unnecessary alternative to the MEGNO for describing the chaotic component. In case of a divided phase space, where regular and chaotic components are mixed, the asymptotically behaviour of MEGNO's final values for regular orbits may hide their different stability levels (see M11 and D12). The SE{\it l}LCE tends to zero for regular orbits. Moreover, the SE{\it l}LCE tends to decrease faster than the LI for regular motion \citep{CS00,CGS03}. So, no matter if the phase space is strongly divided, this behaviour allows the SE{\it l}LCE's final values to show certainly (and faster than the LI) different levels of stability for the regular orbits. Thus, it seems to be appropriate to use the time evolution curves of the MEGNO for the individual analysis of the orbits (M11 and D12) and the final values of the SE{\it l}LCE as a global VIC to obtain the portraits of divided phase spaces (see Sections \ref{FMFT-CIS2}, \ref{STATIONARYS4} and \ref{STARTX0Z0}). 

The MEGNO has a low computational cost and the linear least--squares fitting in the computation of the SE{\it l}LCE does not require a significant amount of extra time. Hence, the SE{\it l}LCE is an alternative quantity that has also a low computational cost.

\subsection{The FLI and the OFLI}\label{CIs&FMFT-S2}
The FLI is a quantity intimately related to the lLCE \citep{FLG97,FGL97,FL00,LF01,GLF02,FL06} and the MEGNO \citep{MCG11}. The FLI is able to distinguish between regular and chaotic (weakly chaotic) motion \citep{FLG97,FGL97} and also between resonant and non--resonant motion \citep{FL00,LF01,GLF02} using only the first part of the computation of the lLCE.

Here, we use the initial definition of the FLI given in \citet{FLG97,FGL97} like in D12, i.e. we use a basis of i.d.v. for its computation. The definition given in \citet{FL00} uses only one i.d.v., the results do not change. Nevertheless, the computing time is obviously reduced \citep{FL00}.

On an $N$--dimensional Hamiltonian $\mathcal{H}$, we follow the time evolution of $2N$ unit i.d.v. (e.g. the ScTS model is $3$--dimensional and the basis consists of 6 i.d.v.). The FLI at time $t$ is defined by the highest norm among the unit i.d.v. of the basis, as follows: 

\[FLI(t)=\sup_t\left[\|\mathbf{w}(t)_1\|,\|\mathbf{w}(t)_2\|,\ldots,\|\mathbf{w}(t)_{2N}\|\right].\] 
For further details, refer to \citet{FLG97,FGL97}.

For both regular and chaotic motion, the FLI tends to infinity as time increases, but with completely different rates. The FLI grows linearly with time for regular motion and it grows exponentially fast for chaotic motion. Furthermore, the FLI grows linearly with time, but with different rates in the case of resonant regular orbits and non--resonant ones. Although the curves corresponding to resonant and non resonant motions are separated, the oscillations due to the distortion of the orbits themselves may hide the distinction between the libration islands and the tori. Therefore, in \citet{FL00} the authors not only introduce the FLI using only one i.d.v. but also replace the definition of the FLI by its running average. We also applied an average to smooth the curves of the FLI.

In the case of the OFLI\footnote{Let us remark that the following definition of the OFLI is slightly different from the definition of the OFLI given in \citet{FLFF02} where the authors use only one i.d.v.}, we take the orthogonal component for the flow for each unit i.d.v. ($w(t)^{\perp}_i,\,i=1,\ldots,2N$) of the basis at every time step and retain the largest component among them: 

\[OFLI(t)=\sup_t\left[w(t)^{\perp}_1,w(t)^{\perp}_2,\ldots,w(t)^{\perp}_{2N}\right].\] 

This modification makes the OFLI a VIC which can easily distinguish periodicity among the orbits of the regular component. The OFLI for periodic orbits oscillates around a constant value, while for quasiperiodic motion and chaotic motion it has the same behaviour as the FLI. 

Refer to Figs. 5 and 7 from \citet{FLFF02} to see the general behaviour of the FLI and the OFLI for regular and chaotic orbits.

\subsection{The APLE}\label{CIs&FMFT-S3}
In the early study of \citet{TPZ97} the authors show that when a nonlinear dynamical system is in the regime of the so--called edge of chaos, the q--entropy rate of increase remains constant for a long time (metastable states). Moreover, they discuss that this behaviour may be associated with a power--law rather than with exponential sensitivity of the orbits to the i.c. The interested reader may refer to \citet{TRLB02} for a ``pedagogical guided tour'' on the subject. 

In \citet{LVE08} the authors find that the time evolution of the i.d.v. for a weakly chaotic orbit $\gamma(\mathbf{x}_0;t)$ (i.e. a combination of a linear and an exponential law: $\mathbf{w}(t)\sim c_\gamma t+e^{d_\gamma t}$, with $c_\gamma\gg 1$ and $d_\gamma\ll 1$) justifies theoretically why metastable states with a constant rate of increase of the q--entropy appear in the nonlinear dynamical system. They argue that this law is reflected upon a transient behaviour in which the growth of the i.d.v. is almost a power law: $\mathbf{w}(t)\propto t^{p_\gamma}$, with $p_\gamma>1$ (and the exponent $p_\gamma$ can be associated to a $q$--exponent via the relationship $p_\gamma=1/(1-q)$). In fact, they introduce a method to compute the $q$--exponent and use it as a VIC (the APLE) in order to distinguish regular orbits from nearby weakly chaotic orbits.    

For an $N$--dimensional Hamiltonian $\mathcal{H}$ consider a partitioning of the $2N$--dimensional phase space $\mathcal{S}$ into a large number of volume elements of size $\delta^{2N}$ for some small $\delta$ and let $\mathbf{x}(0)$ be the initial condition of an orbit located in a particular volume element. The authors introduce the APLE: $p_\gamma$, as follows:

\[
p_\gamma=\frac{\ln\left(\frac{|\mathbf{w}(t)|^2}{|\mathbf{w}(t_1)|^2}\right)}{2\ln\left(\frac{t}{t_1}\right)},
\]
where $|\mathbf{w}(t)|^2=\sum_{k=1}^m\|\mathbf{w}_k(t)\|^2$ and $\mathbf{w}_k(t)$ is one of the $m$ deviation vectors of an orthogonal basis $\{\mathbf{w}_k(t)\}$ of the tangent space to $\mathcal{S}$ at the initial point $\mathbf{x}(0)$. Every $\mathbf{w}_k(t)$ has length greater or equal to $\delta$ and $t_1$ is a transient initial time of orbit evolution.

Therefore, in case of regular motion, $p_\gamma$ tends to 1, while for chaotic motion it grows exponentially fast. Moreover, the APLE shows oscillations around a constant value for a transient time interval before the exponential growth for weakly chaotic orbits. The length of the time interval is directly connected with the hyperbolicity of the local phase space.

Refer to Fig. 2 from \citet{LVE08} to see the general behaviour of the APLE for regular and chaotic orbits. 

\subsection{The FMFT}\label{CIs&FMFT-S4}
The detailed description of the FMFT\footnote{There is a downloadable version of the FMFT in the personal web page of David Nesvorn\'y: \texttt{http://www.boulder.swri.edu/}$\sim$\texttt{davidn/}} has been published on \citet{SN97}. However, we should mention the conceptual aspects of this technique. Since the main advantage of the FMFT is that it improves the computation of the amplitudes and the frequencies given by the FMA \citep{L90,LFC92,L93,L96,PL96,PL98,L03}, we decided to introduce only a short description of the latter. 

Let $\mathcal{H}$ be a non--integrable Hamiltonian which is also non--degenerate:

\[det\left(\frac{\partial\bnu(\mathbf{I})}{\partial\mathbf{I}}\right)=det\left(\frac{\partial^2\mathcal{H}(\mathbf{I})}{\partial \mathbf{I}^2}\right)\neq 0,\]
with $\mathbf{I}\in\mathcal{B}\subset\mathbb{R}^N$ the action-like variables and $\bnu$ the frequency vector. The KAM theory states that for sufficient small values of the perturbation parameter $\epsilon$, there exists a Cantor set $\Omega_\epsilon$ of $\bnu$ that satisfies the Diophantine condition:

\[\|\mathbf{k}\cdot\bnu\|>\frac{\kappa_\epsilon}{|\mathbf{k}|^m},\]
where $\kappa_\epsilon$ and $m$ are constants and $|\mathbf{k}|$ is the norm of $\mathbf{k}$. Furthermore, the system still possesses invariant tori with linear flow (KAM tori). The solutions lie on these invariant tori and are given in complex variables through their Fourier series:

\[z_j(t)=z_{j0}e^{i\nu_j\times t}+\sum_ma_m(\bnu)e^{i\langle \mathbf{m},\bnu\rangle}\]
where the coefficients $a_m(\bnu)$ depend smoothly on the frequencies $\nu_i,\, i=1,\ldots,N$. If we keep all the angle--like variables fixed ($\btheta\in\Pi^N$, $\Pi^N$ is the $N$--dimensional torus, fixed to $\btheta=\btheta_0$), we obtain a frequency map in $\mathcal{B}$ defined as:

\begin{equation}
\mathcal{F}_{\btheta_0}:\mathcal{B}\rightarrow\Omega_\epsilon;\,\,\, \mathbf{I}\rightarrow\mathbf{p}_2(\Psi^{-1}(\btheta_0,\mathbf{I}))
\label{fma-1}
\end{equation}
where $\mathbf{p}_2$ is the projection on $\Omega_\epsilon(\mathbf{p}_2(\mathbf{\phi},\bnu)=\bnu)$. The FMA numerically gives through an iterative process the frequency map $\mathcal{F}$ (i.e. amplitudes and frequencies) defined over the whole domain $\mathcal{B}$, which coincides, within the numerical precision adopted, with the $\mathcal{F}_{\btheta_0}$ of Eq. (\ref{fma-1}) in the set of KAM tori. The frequency map $\mathcal{F}$ is obtained seeking for the quasiperiodic approximations of the solutions over a finite time--span through a finite series of terms:

\begin{equation}
z_j(t)=z_{j0}e^{i\nu_j\times t}+\sum_{k=1}^sa_{m_k}e^{i\langle m_k,\bnu\rangle}.
\label{fma-2}
\end{equation}

Once we obtain the quasiperiodic approximation from Eq. (\ref{fma-2}), we are able to establish a correspondence between the action--like variables ($\mathbf{I}\in\mathcal{B}$) and the rotation numbers (or frequency vector $\bnu\in\Omega_\epsilon$). This correspondence is the so--called FMA \citep{L90}. Regular regions of the phase space allow a very accurate estimation of real rotation numbers in a given time interval (for instance, $\propto 1/T^4$ using the MFT with the Hanning window, $\propto 1/T^2$ using the MFT without the Hanning window, $\propto 1/T$ using the usual FFT techniques, where $[-T:T]$ is set as the whole time--span), and should not differ from the rotation numbers obtained in another time interval, considering a certain accuracy. Strictly speaking, the frequencies are only defined on KAM tori, the FMA numerically estimates over a finite time--span a frequency vector for any initial condition, though. If the estimates of the rotation numbers given by the FMA vary (above the required accuracy) between different time periods, it implies that the corresponding KAM tori are destroyed. Thus, the study of this frequency map $\mathcal{F}$ (i.e. the FMA) regarding its constancy in time provides important clues about the dynamics of the system. 

For a comprenhensive discussion of the FMA refer to \citet{L03}.

In order to calculate the frequencies with the FMFT, we need to compute the equations of motion in the ScTS model. We use the Taylor method \citep{JZ05} that proved to be very convenient for the task in D12. The precision in the computation of the coordinates is $10^{-15}$. The numerical computation of the VICs was done using the DOPRI8 routine \citep{PD81}. This routine achieved better results than Taylor when we required the simultaneous integration of the coupled system of equations of motion and variational equations in the ScTS model (see also D12). The conservation of the energy with the DOPRI8 routine was $\sim10^{-13},\,10^{-14}$.

All the computations in the forthcoming investigation were done using the following configuration. Hardware: CPU, 2 x Dual XEON 5450, Dual Core 3.00GHz; M.B., Intel S5000VSA; RAM, 4GB(4x1GB), Kingston DDR--2, 667MHz, Dual Channel. Software: gfortran 4.2.3. 

\section{The FMFT as a global indicator of chaos}\label{theFMFT}
In this section, we compare the performances of the FMFT and the SE{\it l}LCE as global indicators of chaos. 

\subsection{Accuracy in the estimation of the frequencies}\label{FMFT-CIS1} 
In order to analyse the nature of each orbit we apply the FMFT to the following function in each degree of freedom (hereinafter, d.o.f.):

\begin{equation}
\psi_j(t_k)=\psi^{\mathbb{R}}_j(t_k)+i\psi^{\mathbb{C}}_j(t_k)
\label{fmft-ci1}
\end{equation}
with $\psi_j$ ($j=x,y,z$) a complex function, where $\psi^{\mathbb{R}}_j$ (the real part of $\psi_j$) corresponds to the Cartesian coordinates ($x,y$ and $z$) and $\psi^{\mathbb{C}}_j$ (the conjugate part of $\psi_j$) corresponds to the Cartesian velocities ($v_x,v_y$ and $v_z$). The $t_k$ with $k=1,\ldots,n$ are the trajectory samplings. Every set of points $\psi_j$ generates a quasiperiodic approximation and we keep the frequency $\nu_j(\neq 0)$ with the maximum amplitude. We call this frequency the fundamental frequency of the system in the $j$ d.o.f. 

Theoretically speaking, if the fundamental frequencies correspond to a regular orbit, they should remain constant in time. Numerically speaking, the variation exists and it is due to the precision of the computation. Thus, in order to determine such a precision for our particular experiment, we take samples of regular orbits, according to both convergent values of the LI and the MEGNO ($<\ln(t)/t$ or $<2.01$, respectively, see \citet{MGC11}).
 
We are aware that samples obtained from the use of VICs can only be statistically rich in regular orbits. That is, the actual values of the VICs are reached only at infinite time and as we have just truncated approximations of those values, some of the orbits might be missclassified. In order to be as certain as possible about the dominant type of orbits from the samples, we follow their behaviour for $10^3\,T_c(E)$ (see Section \ref{CIs&FMFT-S1}). In a galaxy like the one represented by the ScTS model, the crossing time is $0.5$ u.t. If we suppose that the Hubble time is of the order of $10^3$ crossing times (a high value indeed), then we are integrating the orbits for $\sim 234$ and $\sim 14$ Hubble times for the energy surfaces $-0.1$ ($T_c(-0.1)=1.17\times 10^2$ u.t.) and $-0.7$, respectively. Therefore, $10^3\,T_c(E)$ seems to be a lapse of time large enough to provide a physical meaningful characterization of the motion of the orbits and reliable convergent values of the VICs for most of the orbits in the sample. Finally, for all practical purposes, most of the orbits of the samples are regular orbits.

\begin{figure*}
\begin{tabular}{cc}
\resizebox{75mm}{!}{\includegraphics{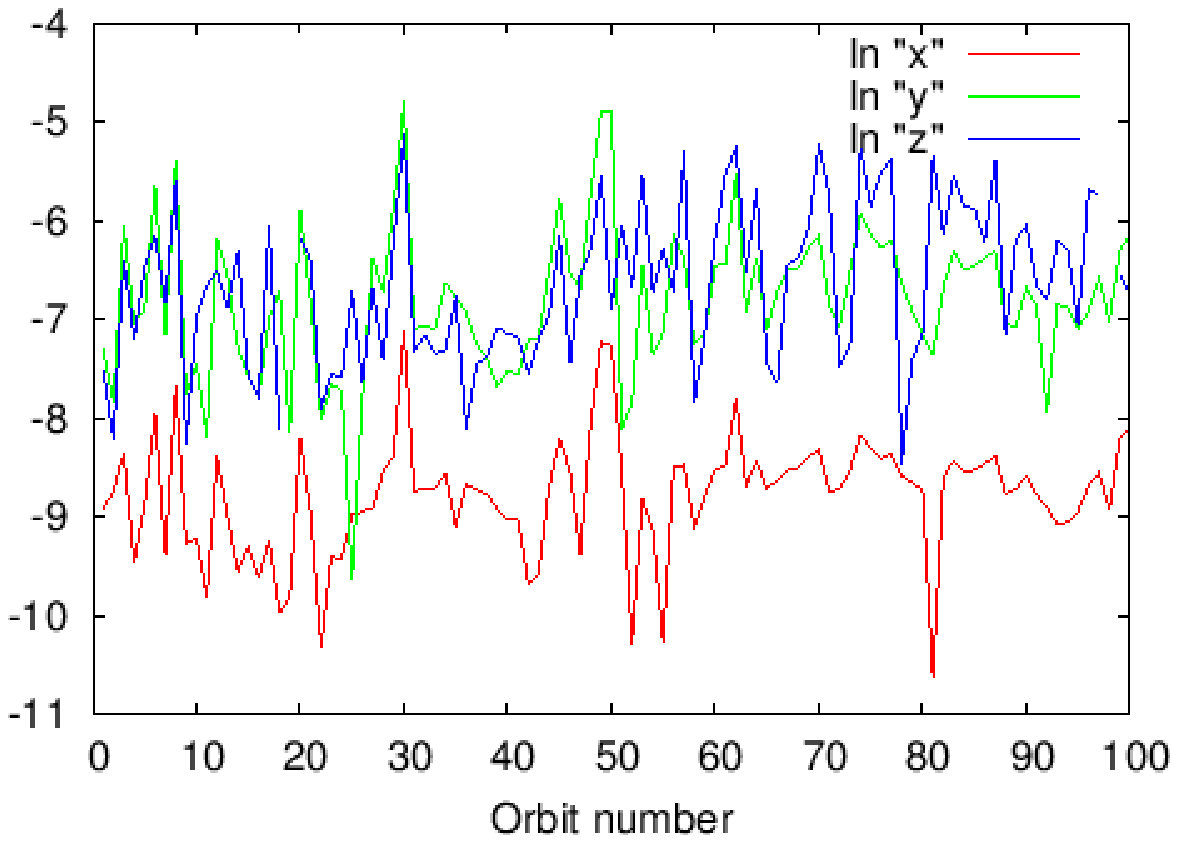}}& 
\resizebox{75mm}{!}{\includegraphics{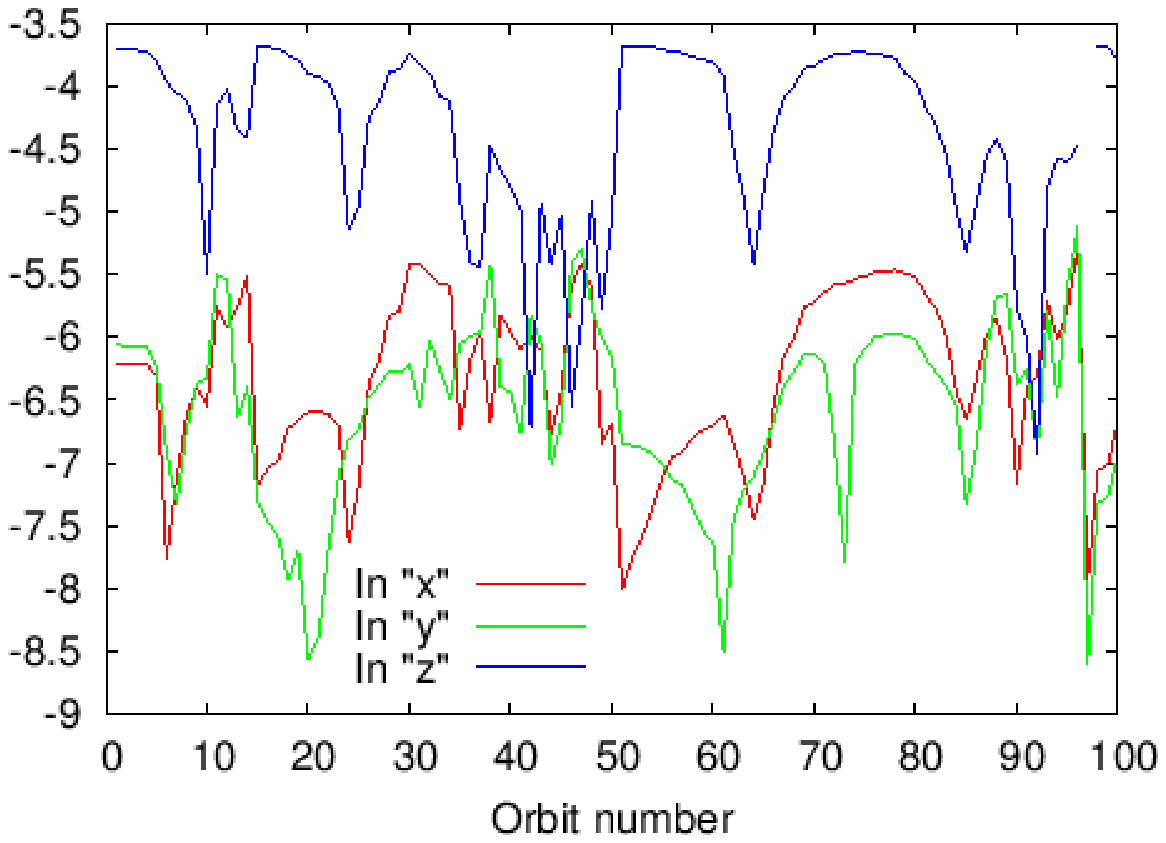}}\\
\resizebox{75mm}{!}{\includegraphics{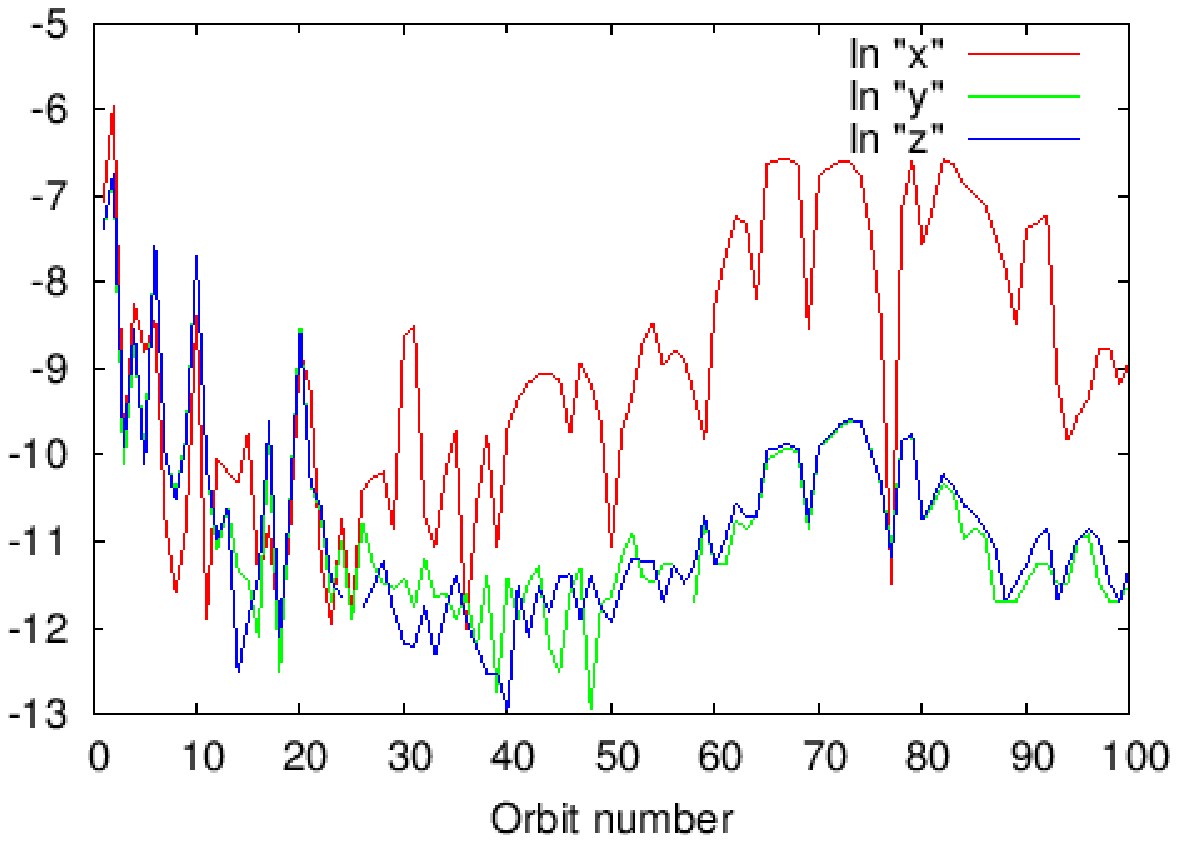}}& 
\resizebox{75mm}{!}{\includegraphics{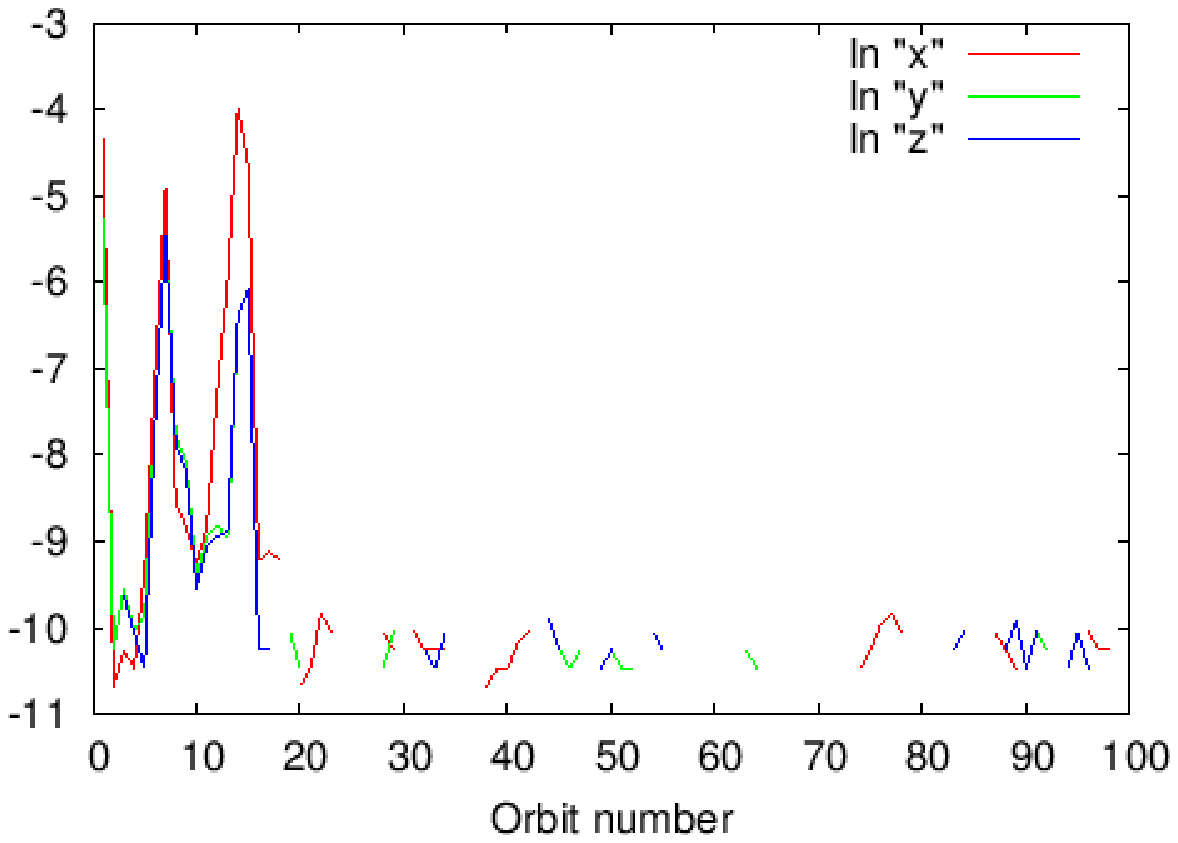}}
\end{tabular}
\caption{Variations of the fundamental frequencies (in logarithmic scale) computed with the FMFT for four samples of 100 regular orbits on each d.o.f.: on the $p_{x_0}-p_{z_0}$ start space and on the energy surfaces $-0.1$ (top left panel) and $-0.7$ (top right panel), and  on the $x_0-z_0$ start space and on the energy surfaces $-0.1$ (bottom left panel) and $-0.7$ (bottom right panel). The missing points are due to the computational precision.}
\label{FMFT-CI-1}
\end{figure*}

The samples are taken in the $p_{x_0}-p_{z_0}$ and the $x_0-z_0$ start spaces for four different energy surfaces, namely: $-0.1$, $-0.3$ ($T_c(-0.3)=24$ u.t.), $-0.5$ ($T_c(-0.5)=12$ u.t.) and $-0.7$ (i.e. a total of eight samples). There are 100 i.c. in each sample and we integrate the equations of motion for $4.5\times 10^2\,T_c(E)$ for every i.c. Although $10^3\,T_c(E)$ is the time--span we use to compute the VICs reliably, the results in the determination of the fundamental frequencies with the FMFT are not very different using only $3\times 10^2\,T_c(E)$. Besides, the CPU time is certainly reduced. Therefore, we apply the FMFT on two $50\%$ overlapping time intervals of $3\times 10^2\,T_c(E)$ \citep{WF98}\footnote{The time intervals are $[0:3\times 10^2\,T_c(E)]$ and $[1.5\times 10^2\,T_c(E):4.5\times 10^2\,T_c(E)]$.}. The number of points used in the estimation of the frequencies is $8192$ which makes $\sim27$ points for each $T_c(E)$. Finally, we estimate the fundamental frequencies $\nu_j^{(i)}$ for each time interval and calculate the differences (the accuracy in the determination of the frequencies should be $\sim10^{-9}$ according to Section \ref{CIs&FMFT-S4} and \citet{M06}). The concomitant results for the energy surfaces $-0.1$ and $-0.7$ for both the $p_{x_0}-p_{z_0}$ and the $x_0-z_0$ start spaces are shown in Fig. \ref{FMFT-CI-1}. 

In Fig. \ref{FMFT-CI-1} we show that the differences in the computed values of the fundamental frequencies between both overlapping time intervals for regular orbits (actually, some of them might be chaotic) is not generally below the theoretical value, particularly in the $p_{x_0}-p_{z_0}$ start space. Thus, there exists a range of values where the regular orbits could not be reliably distinguished from the chaotic orbits by means of the computation of the fundamental frequencies with the FMFT.

In Table \ref{tablefmft} we show the corresponding arithmetic means and standard deviations of the variations of the fundamental frequencies for the energy surfaces: $-0.1$, $-0.3$, $-0.5$ and $-0.7$ on the $p_{x_0}-p_{z_0}$ start space. The arithmetic means show an average conservation between the 7th and the 8th decimal in the fundamental frequencies. Nevertheless, the standard deviations oscillate around an average of the 5th decimal. Therefore, there is an interval where the FMFT can not distinguish clearly between regular and chaotic orbits in the experiment. However, we can infer certain limits which can help us to identify if the orbit is regular or chaotic using the FMFT. For instance, the variations in the computed fundamental frequencies for the regular orbits only affect decimals higher than the 4th (see Fig. \ref{FMFT-CI-1}). Thus, if there is a difference in the computation of the fundamental frequencies of the orbit in decimals lower than the 4th, it is highly probable that the orbit is chaotic. This result is in complete agreement with the one shown in \citet{M06} for the same ScTS model and with the practical limit used also in \citet{AMNZ07,MNZ09}.

\begin{table}
\centering
\caption{Arithmetic means and standard deviations of the differences on the fundamental frequencies between both overlapping time intervals for the samples of regular orbits used in Fig. \ref{FMFT-CI-1}. The values are ordered by energy surface and d.o.f.}
\begin{tabular}{@{}cccc@{}}
\hline
Energy $-0.1$  & $x$ & $y$ & $z$\\ 
\hline 
 Arithmetic means & $\sim1.2^{-9}$ & $\sim-2.3^{-7}$ & $\sim-3.4^{-7}$\\
 Standard deviations & $\sim1.2^{-7}$ & $\sim2.5^{-5}$ & $\sim1.7^{-5}$\\
\hline
Energy $-0.3$   & & & \\
\hline 
 Arithmetic means & $\sim-1.8^{-8}$ & $\sim1.2^{-6}$ & $\sim5.9^{-6}$\\ 
 Standard deviations & $\sim4.6^{-6}$ & $\sim4.3^{-4}$ & $\sim5.9^{-4}$\\
\hline
Energy $-0.5$   & & & \\
\hline 
 Arithmetic means & $\sim5.3^{-9}$ & $\sim-1.9^{-8}$ & $\sim1.1^{-7}$ \\ 
 Standard deviations & $\sim5.3^{-7}$ & $\sim2.7^{-6}$ & $\sim1.3^{-5}$\\
\hline
Energy $-0.7$   & & &  \\
\hline 
 Arithmetic means & $\sim3.5^{-7}$ & $\sim-6.9^{-8}$ & $\sim-1.9^{-5}$\\ 
 Standard deviations & $\sim1.6^{-5}$ & $\sim1.3^{-5}$ & $\sim1.2^{-3}$\\
\hline
\end{tabular}
\label{tablefmft}
\end{table}

The results for the $x_0-z_0$ start space are similar than those for the $p_{x_0}-p_{z_0}$ start space. Nevertheless, for lower energy values (Fig. \ref{FMFT-CI-1}, bottom right panel), the averaged precision in the computation of the frequencies improved substantially.

We also make some experiments with the FMFT, using synthetic signals for verifying the precision in the computation of the fundamental frequencies. The results were remarkable better and are closer to the theoretical estimates \citep{M06}. Nevertheless, this is not the case for the relatively complex ScTS model and the precision in the estimations is clearly diminished. Thus, we use the same model and the same spaces we study to calibrate the FMFT.

\subsection{A comparative evaluation between the FMFT and the SE{\it l}LCE}\label{FMFT-CIS2}
\begin{figure}
\begin{tabular}{c}
\resizebox{75mm}{!}{\includegraphics{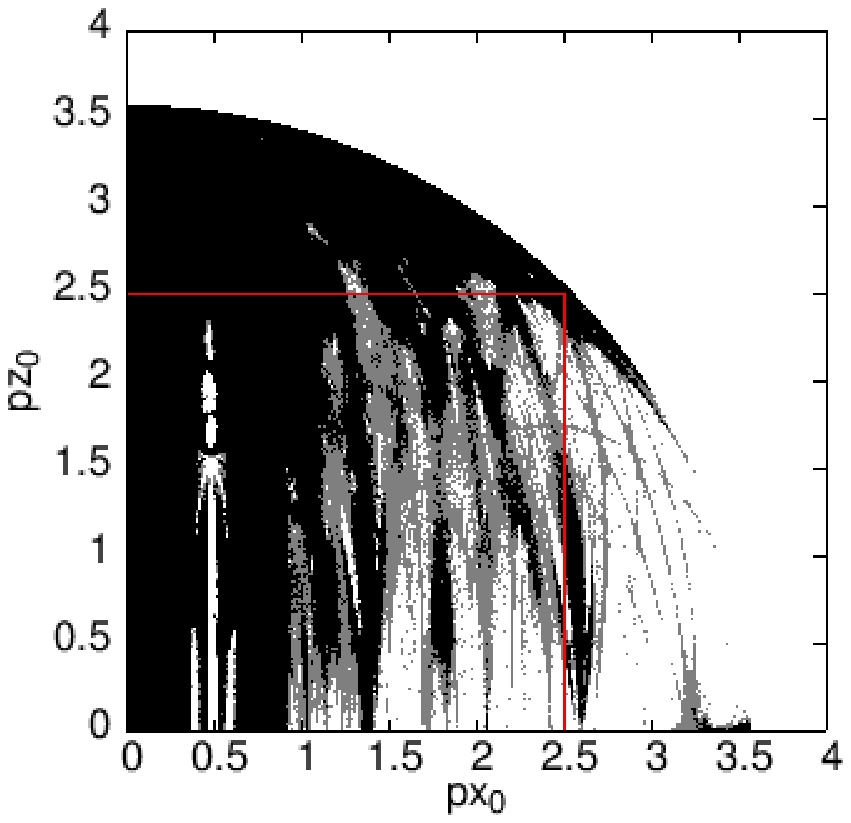}}\\ 
\resizebox{75mm}{!}{\includegraphics{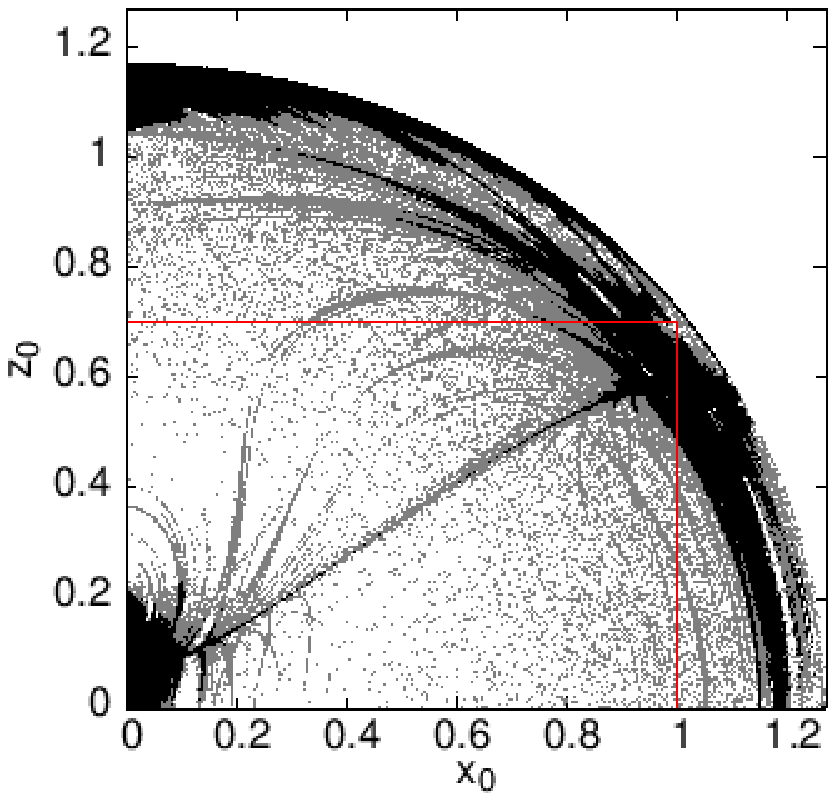}}
\end{tabular}
\caption{On the top panel we show in a red square the region we will study in the $p_{x_0}-p_{z_0}$ start space and on the bottom panel, the one we will study in the $x_0-z_0$ start space. Both regions are located in the energy surface defined by the constant value $-0.7$. We take a total integration time of $7\times10^3$ u.t. The phase space portraits are generated in grey--scale with the MEGNO. Values of the MEGNO below the threshold $2.01$, i.e., regular orbits, are depicted in white. In grey we enclosed the values of the MEGNO in the interval $(2.01:20)$, which means moderate to strong chaos. In black, we show values of the MEGNO above $20$, i.e. regions with very strong chaos.}
\label{regiones-estacionario-inicializacion}
\end{figure}

We are interested in testing the performance of the FMFT as a global indicator of chaos. Thus, we apply it together with a variational indicator in order to characterise the phase space portraits of two different regions. One of the regions is located in the $p_{x_0}-p_{z_0}$ start space (the region is identified in the top panel of Fig. \ref{regiones-estacionario-inicializacion}), whereas the other is located in the $x_0-z_0$ start space (which is identified in the bottom panel of the same figure). Both regions belong to the energy surface $-0.7$. The VIC selected for this study is the SE\textit{l}LCE (see Section \ref{CIs&FMFT-S1}). 

In order to apply the FMFT as a global indicator of chaos, we follow the same procedure as described in Section \ref{FMFT-CIS1}. Thus, we consider two overlapping time intervals of $10^3$ periods. The first time interval is $[0:7\times 10^3]$ u.t. and the second time interval is $[3.5\times 10^3:10.5\times 10^3]$ u.t. According to Section \ref{CIs&FMFT-S4}, the accuracy in the determination of the frequencies should be $\sim10^{-11}$. The computation is over 624100 i.c., within the $p_{x_0}-p_{z_0}$ start space, covering the region depicted on the top panel of Fig. \ref{regiones-estacionario-inicializacion} and over 596258 i.c., within the $x_0-z_0$ start space, covering the region depicted on the bottom panel of the same figure. 

To use the FMFT as a global indicator of chaos and compute the corresponding phase space portraits, we need to define a new quantity \citep{WF98}. Such a quantity is $\log(\Delta F)$, $\Delta F$ being $|\nu_x^{(1)}-\nu_x^{(2)}|+|\nu_y^{(1)}-\nu_y^{(2)}|+|\nu_z^{(1)}-\nu_z^{(2)}|$, where $\nu_j^{(i)}$ is the frequency (computed with the FMFT) associated with the d.o.f ``j'' for the time interval ``(i)''. All the fundamental frequencies have to be computed in both intervals. Finally, the phase space portraits shown by the quantity $\log(\Delta F)$ consist of 622521 i.c. for the $p_{x_0}-p_{z_0}$ start space and 594690 i.c. for the $x_0-z_0$ start space. 

\begin{figure}
\begin{tabular}{c}
\resizebox{75mm}{!}{\includegraphics{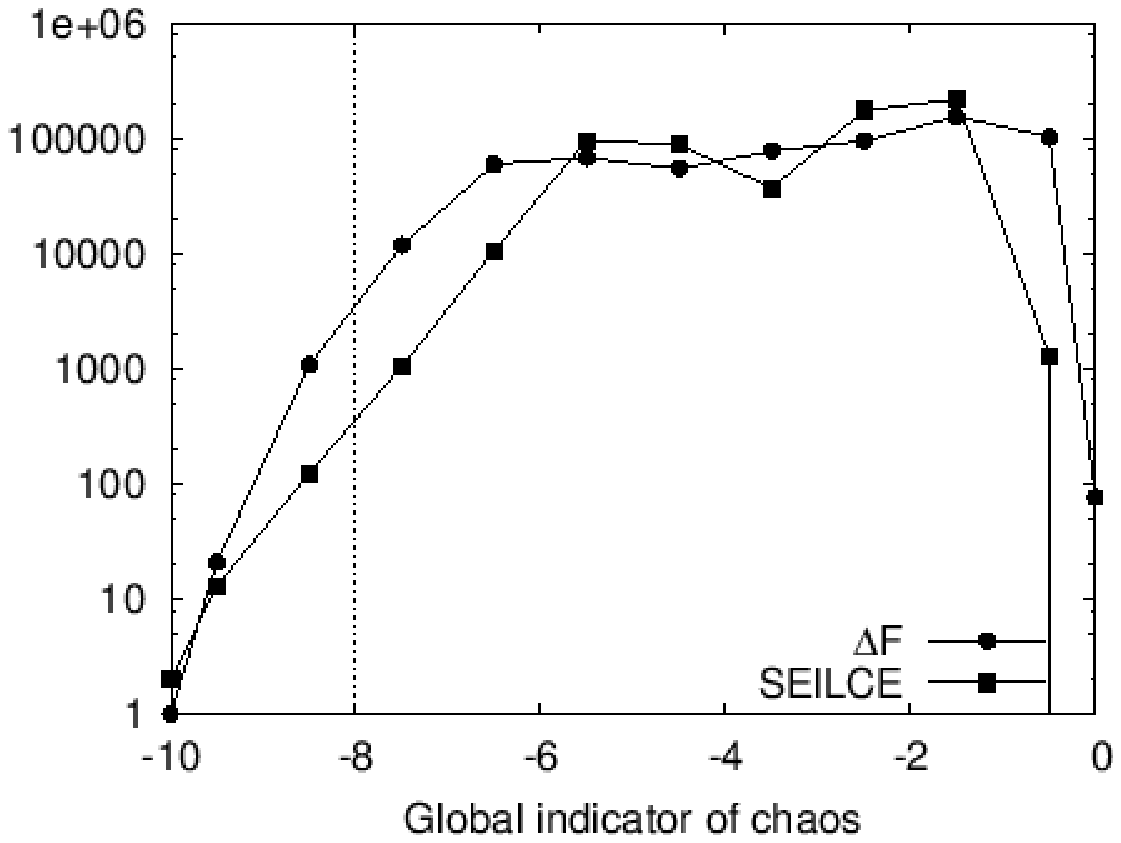}}\\ 
\resizebox{75mm}{!}{\includegraphics{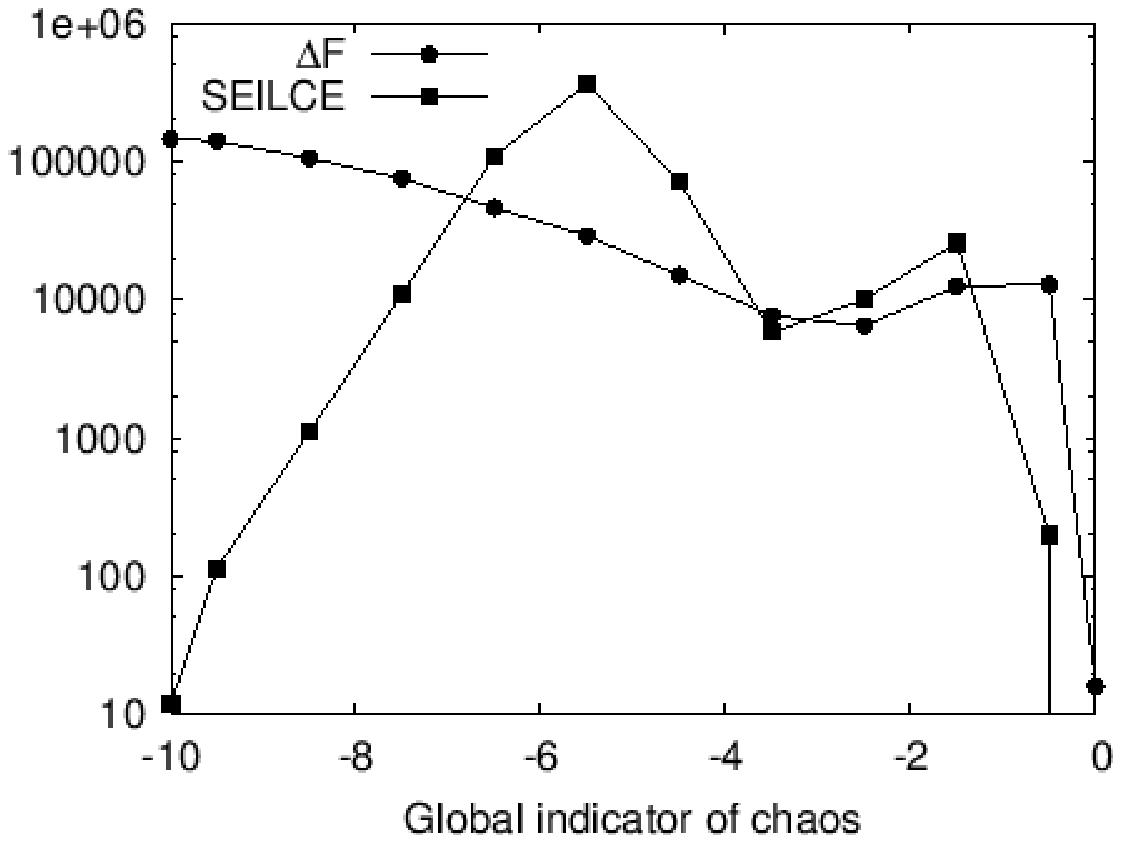}}
\end{tabular}
\caption{Histograms in logarithmic scale showing the bimodal distribution of both global indicators of chaos on the $p_{x_0}-p_{z_0}$ (top panel) and the $x_0-z_0$ (bottom panel) start spaces on the energy surface $-0.7$. The cut-off level of $10^{-8}$ is depicted on the top panel.}
\label{add-histograms}
\end{figure}

In order to make appropriate comparisons on each start space, we need to evaluate the $\log(\Delta F)$ and the SE\textit{l}LCE on equal standings. Thus, it is necessary to decide a cut--off level for the distributions of their values to use the same scales for both indicators. 

\begin{figure*}
\begin{tabular}{cc}
\resizebox{75mm}{!}{\includegraphics{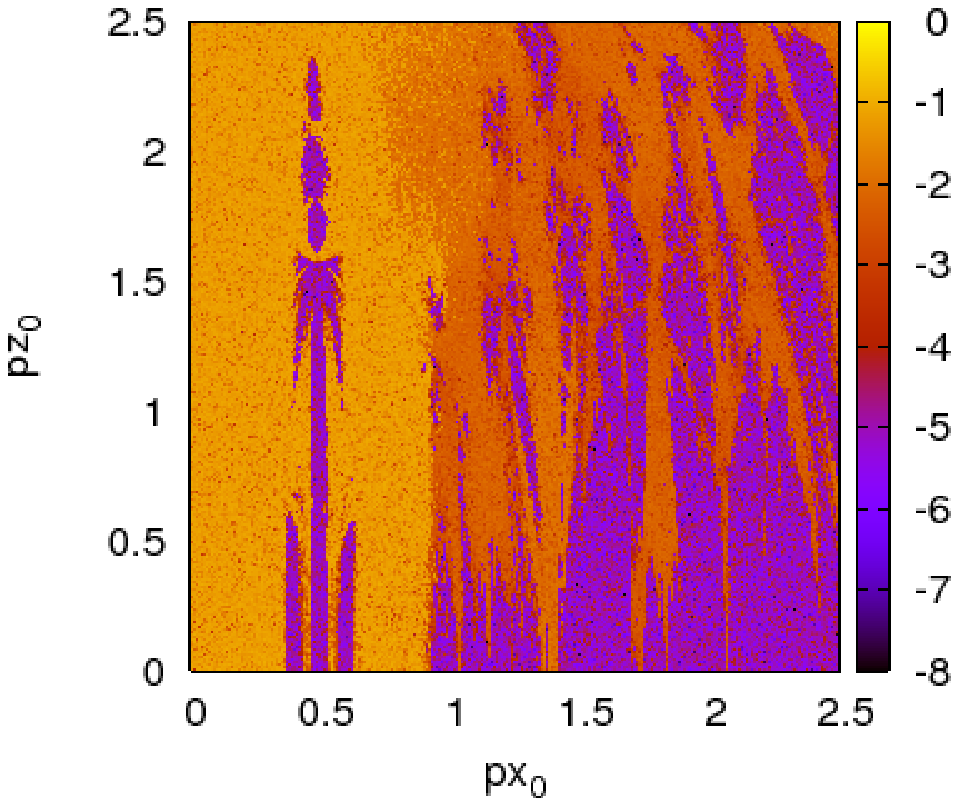}}& 
\resizebox{75mm}{!}{\includegraphics{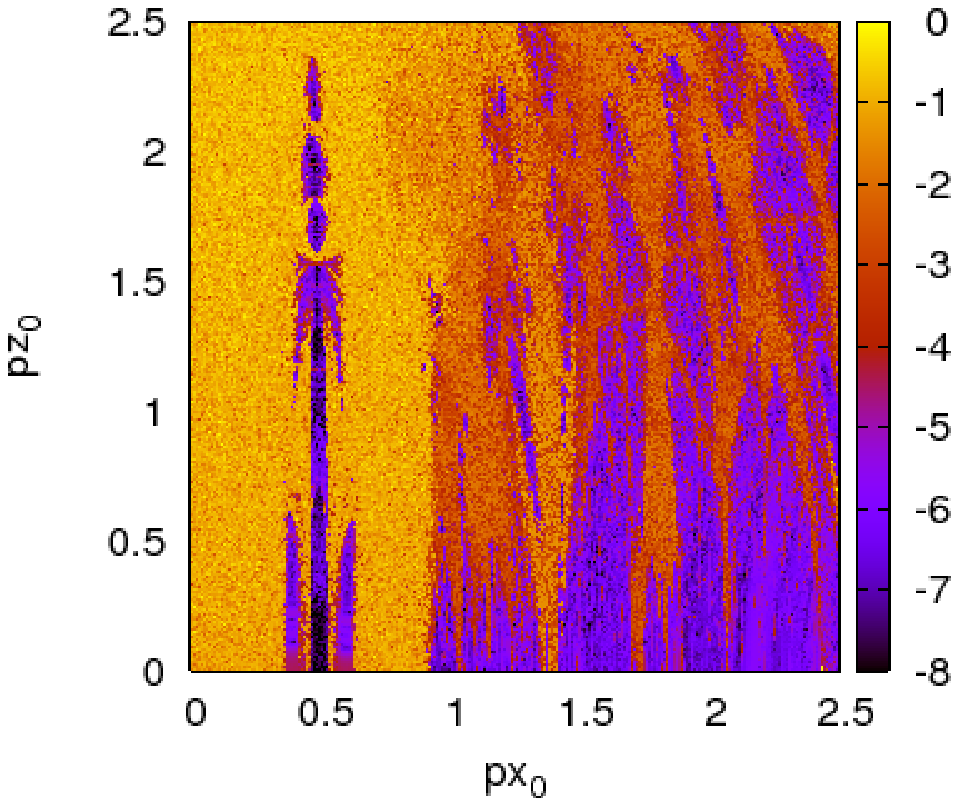}}\\
\resizebox{75mm}{!}{\includegraphics{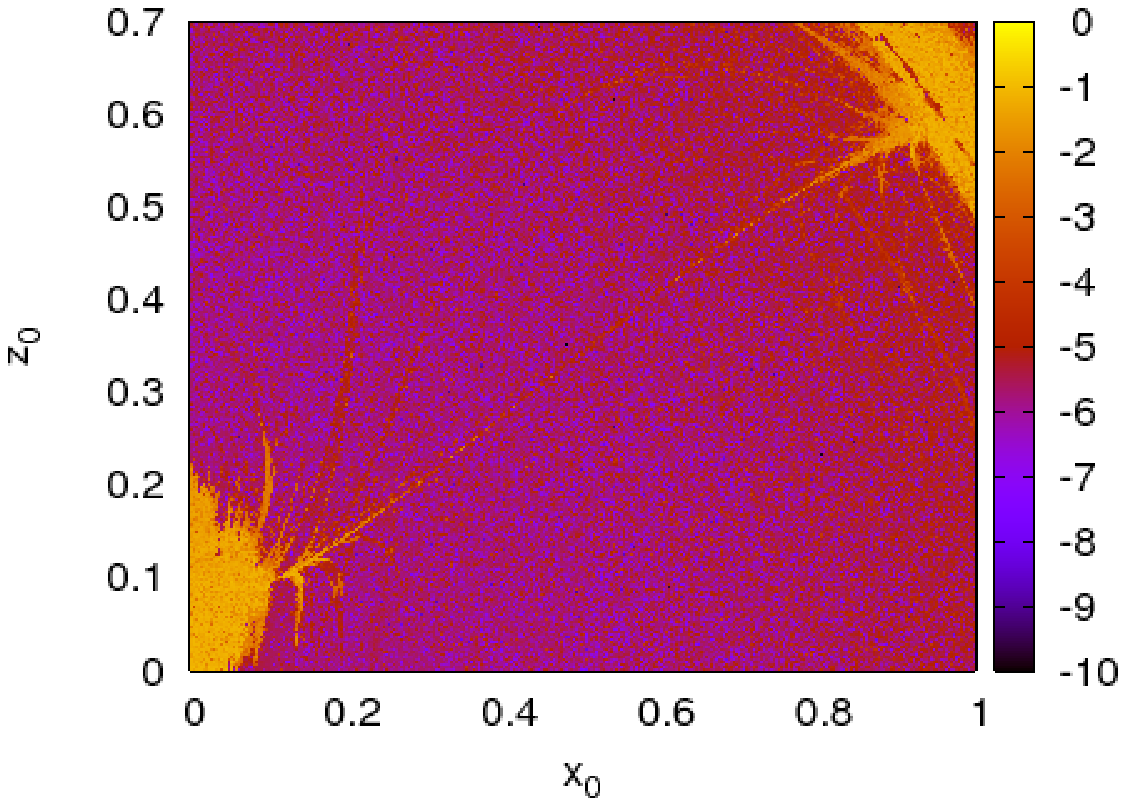}}& 
\resizebox{75mm}{!}{\includegraphics{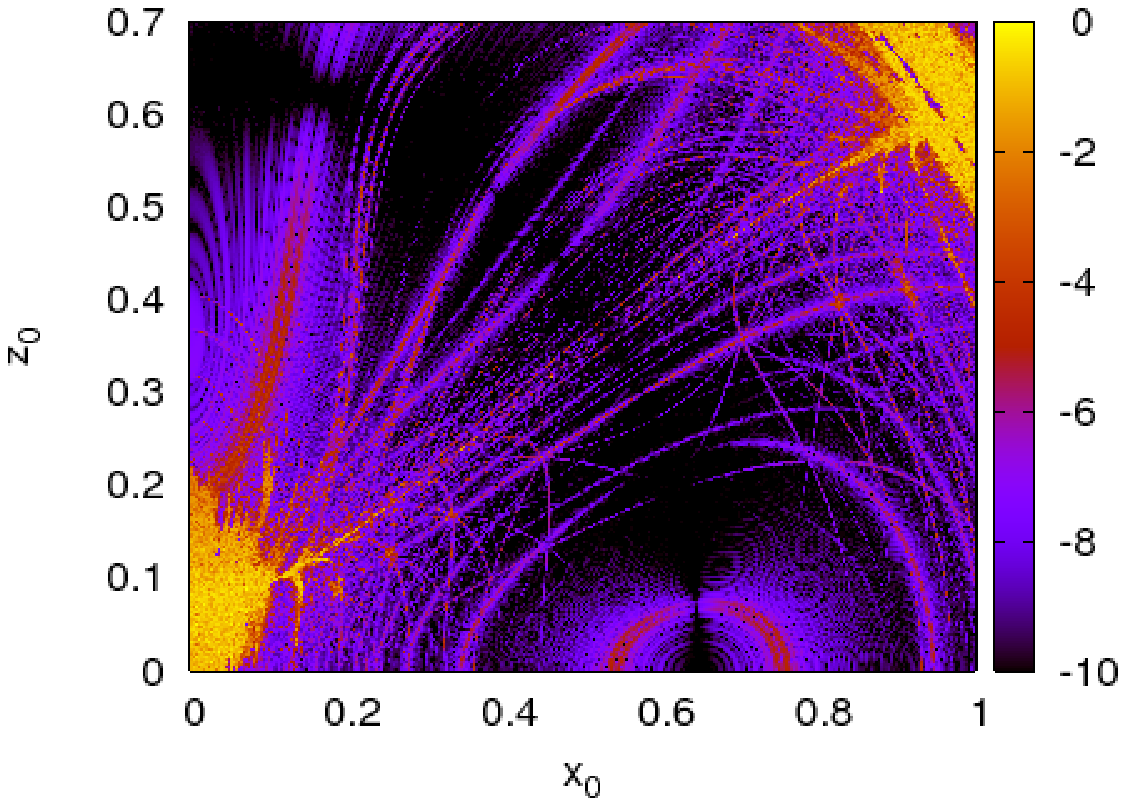}}
\end{tabular}
\caption{Phase space portraits corresponding to both regions within the $p_{x_0}-p_{z_0}$ (top panels) and the $x_0-z_0$ start (bottom panels) spaces on the energy surface $-0.7$. The samples used to compute the SE\textit{l}LCE consist of 624100 and 596258 (top and bottom left panels, respectively) i.c. and $7\times 10^3$ u.t. The samples used to compute the $\log(\Delta F)$ consist of 622521 and 594690 (top and bottom right panels, respectively) i.c. and two overlapping time intervals of $7\times 10^3$ u.t. each. Chaotic regions are identified by the values of the indicators that are higher than $\sim-4,\,\sim-3$ (notice that the values of the indicators are in logarithmic scale). See text for further details.}
\label{FMFT-CIS1-1}
\end{figure*}

In Fig. \ref{add-histograms} we show the histograms according to the final values of the techniques. The histograms on the top panel correspond to the $p_{x_0}-p_{z_0}$ start space and the histograms on the bottom panel correspond to the $x_0-z_0$ start space. A clear bimodal distribution (one peak corresponding to regular orbits, and the other to chaotic orbits) is shown for both indicators. In case of the $p_{x_0}-p_{z_0}$ start space, the indicators have less than $1\%$ values for the orbits below $10^{-8}$ (pointed out on the top panel of the same figure). Thus, the scales on the $p_{x_0}-p_{z_0}$ start space go from $10^{-8}$ to $0$. The similarities between both histograms on the $p_{x_0}-p_{z_0}$ start space make easier the selection of a common cut-off level. However, there are important differences between the histograms on the $x_0-z_0$ start space. The SE\textit{l}LCE shows a bimodal distribution within the interval $[\sim10^{-10}:0)$ with its peaks well separated. The peaks in the distribution on the $x_0-z_0$ start space match the ones found for the same indicator, but on the $p_{x_0}-p_{z_0}$ start space (see both panels in Fig. \ref{add-histograms}). This is not the case for the $\log(\Delta F)$. The peak that corresponds to the regular component is beyond the value $10^{-10}$ (see Fig. \ref{add-histograms}, bottom panel and Fig. \ref{FMFT-CI-1}, bottom right panel). Nevertheless, we find that the number of orbits with values of the SE\textit{l}LCE below the value $10^{-10}$ are negligible and the $\log(\Delta F)$ does not show further structure below that value. Thus, we decided to use that value as the cut--off level in the distributions on the $x_0-z_0$ start space to build the scales.  

Notice also that, according to $\log(\Delta F)$ on the $p_{x_0}-p_{z_0}$ start space, the peak corresponding to regular orbits is $\sim3.1\times10^{-6}$. This value confirms the conclusions drawn in Section \ref{FMFT-CIS1} with a sample composed only of 100 orbits for the same start space. That is, if we use the arithmetic means for the energy surface $-0.7$ (Table \ref{tablefmft}), we arrive to a similar value of $\sim1.9\times10^{-5}$. Furthermore, the minimum between both peaks (which gives a clear distinction between the regular and the chaotic component) in the $p_{x_0}-p_{z_0}$ start space is taken around the value $10^{-4}$. This value also matches the one considered to separate regular from chaotic orbits for the same start space in Section \ref{FMFT-CIS1}. 

In Fig. \ref{FMFT-CIS1-1} we present the phase space portraits given by the SE\textit{l}LCE (left panels) and the $\log(\Delta F)$ (right panels) for both selected regions in the $p_{x_0}-p_{z_0}$ (top panels) and the $x_0-z_0$ (bottom panels) start spaces. 

The phase space portraits of the region inside the $p_{x_0}-p_{z_0}$ start space given by the SE\textit{l}LCE and the $\log(\Delta F)$ present the same structures. In spite of this high level of coincidence in the description of the $p_{x_0}-p_{z_0}$ start space, there are differences in the $x_0-z_0$ start space. On the bottom right panel of Fig. \ref{FMFT-CIS1-1}, the quantity $\log(\Delta F)$ shows an amount of spurious structures\footnote{Some of them, are due to the phenomenon of Moir\'e, which is inherent in discrete Fourier transform techniques, \citet{BBB09}.} not shown by the variational indicator (bottom left panel of the same figure). In fact, the spurious structures make difficult to select a threshold to distinguish between regular and chaotic orbits. Still, it is possible to define a cut-off value for the $\log(\Delta F)$, but the procedure is more handmade. For example, if we take as regular orbits those preserving $\gtrsim 4$ decimals in their fundamental frequencies (see Section \ref{FMFT-CIS1} and the previous discussion), we recover the portrait given by the SE\textit{l}LCE for the $x_0-z_0$ start space. However, the SE\textit{l}LCE (as well as most of the VICs studied in previous papers, for instance M11 and D12) shows a large separation of the different kinds of motion in both portraits (see also Fig. \ref{add-histograms}), and the choice of a threshold (if it is not already defined) is easier.

These results, which are not as clear as those given by the variational indicators in terms of motion separation, make the FMFT a second choice to study the global dynamics in a divided phase space. 

The procedure used to determine the chaoticity or regularity of the orbits with the FMFT is standard. The deficiency in the descriptions is basically due to the method's sensitivity to its parameters. Therefore, the reliability of the FMFT as a global indicator of chaos is, comparatively speaking, limited.  

\subsubsection{The computing times}\label{FMFT-CIS2S1} 
The VICs need to compute the equations of motion together with the variational equations to determine the nature of the orbits. This process may take a large amount of time. On the other hand, the FMFT works with the computation of the frequencies, which is a fast process, but the determination of the $\log(\Delta F)$ to characterise the nature of the orbits involves further calculations. For example, the computation of the SE\textit{l}LCE (a fast VIC) to obtain the top and bottom left panels of Fig. \ref{FMFT-CIS1-1} took $\sim$670 hs and $\sim$330 hs, respectively, by using the DOPRI8 routine. Instead, the computation of the fundamental frequencies (nothing more than the the line of biggest amplitude in each d.o.f. was computed) with the FMFT for 624100 and 594690 i.c. took $\sim$30 hs for both experiments, after the integration of the equations of motion. The time spent in this integration for $7\times 10^3$ u.t. was of $\sim$570 hs and $\sim$280 hs for the regions in the $p_{x_0}-p_{z_0}$ and the $x_0-z_0$ start spaces, respectively. In both cases we used the Taylor method (see Section \ref{CIs&FMFT-S4}). Furthermore, if we want to use the FMFT as a global indicator of chaos, we need to compute the $\log(\Delta F)$, i.e. we need to compute the frequency vector for two time intervals. If we overlap the time intervals in a 50\%, we economize on computing time. The total time interval must be $1.05\times 10^4$ u.t. under these circumstances. That is, 50\% larger than the one used by the variational indicators. Finally, the time required by the FMFT extends from $\sim$600 ($\sim$570 + $\sim$30 hs) to $\sim$885 hs and from $\sim$310 ($\sim$280 + $\sim$30 hs) to $\sim$450 hs to obtain the portraits on the top and bottom right panels of Fig. \ref{FMFT-CIS1-1}, respectively.

The advantage of the fast computation of the frequencies is certainly lost in the time--consuming process involved in the computation of the $\log(\Delta F)$.

In conclusion, the FMFT as a global indicator of chaos seems to be rather inconvenient when there are other alternatives such as fast VICs like the SE\textit{l}LCE. 

\section{The selection of proper VICs}\label{STATIONARY}
The aim of this section is to select the proper VICs to be used in the forthcoming experiments on the $p_{x_0}-p_{z_0}$ and the $x_0-z_0$ start spaces of the ScTS model (Section \ref{thespaces}). 

\begin{figure}
\begin{tabular}{c}
\resizebox{75mm}{!}{\includegraphics{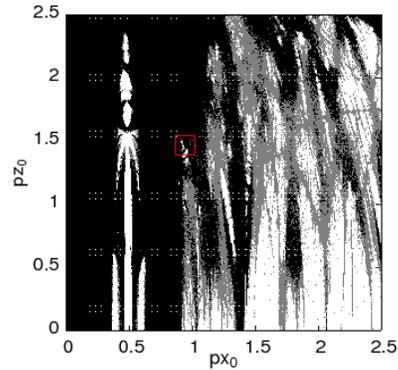}}
\end{tabular}
\caption{We signal with a red square the region in the $p_{x_0}-p_{z_0}$ start space (energy: $-0.7$) under analysis. The grey--scale portrait is obtained with the final values of the MEGNO. In white, we identify the regular orbits and thus the MEGNO values $\lesssim2.01$. In grey, we identify regions of moderate to strong chaos, that is MEGNO values in the interval $(2.01:20)$. In black, with MEGNO values $\gtrsim20$, we identify very strong chaos.}
\label{region-in-stationary}
\end{figure}

We first study a small region in the $p_{x_0}-p_{z_0}$ start space (within the intervals $p_{x_0}\in[0.87:1.02]$ and $p_{z_0}\in[1.39:1.54]$) on the energy surface $-0.7$, shown in the red square in Fig. \ref{region-in-stationary}). We use the MEGNO, the SE\textit{l}LCE, the FLI, the OFLI and the APLE. This first experiment provides us with a comparison between quite new techniques (the SE\textit{l}LCE and the APLE) and VICs with excellent performances shown in previous studies (the MEGNO and the FLI/OFLI, e.g. in M11 and D12). 

In addition, we use the same $10^3$ characteristic times along the experiments.

We are interested in the resolving power and the speed of convergence of the VICs. Therefore, we compute 10201 i.c. within the region marked in Fig. \ref{region-in-stationary} for $7\times 10^3$ u.t. 

\subsection{The resolving power of the VICs}\label{STATIONARYS1} 
First, we briefly discuss the resolving power of the VICs under the circumstances of the experiment. 

As we are looking for VICs to study big samples of i.c., we consider the final values of the indicators rather than their time evolution curves. Furthermore, when the indicator grows or decreases exponentially, it is convenient to stop the integration process at a particular saturation value. Thus, instead of simply registering the final values of the indicator, we can record the time needed to reach such saturation value, i.e. the saturation times \citep{SBA07}. The final values together with the saturation times of the VIC should be considered for the description of the phase space portrait (M11 and D12).    

\begin{figure*}
\begin{tabular}{cc}
\resizebox{75mm}{!}{\includegraphics{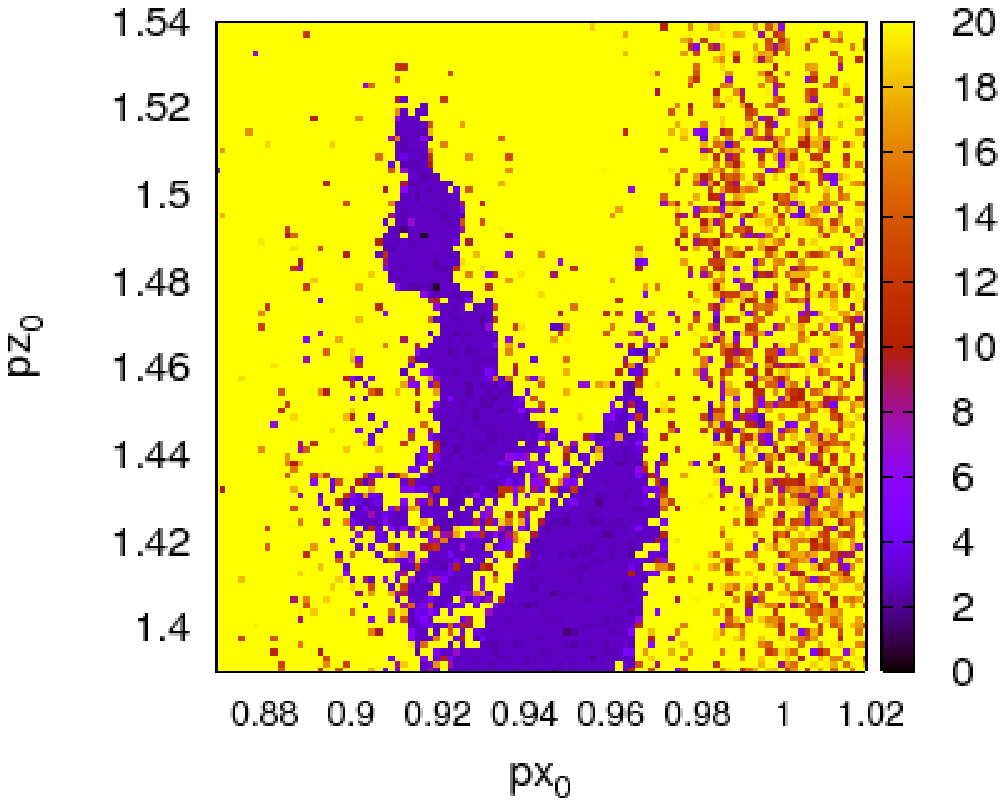}}&
\resizebox{75mm}{!}{\includegraphics{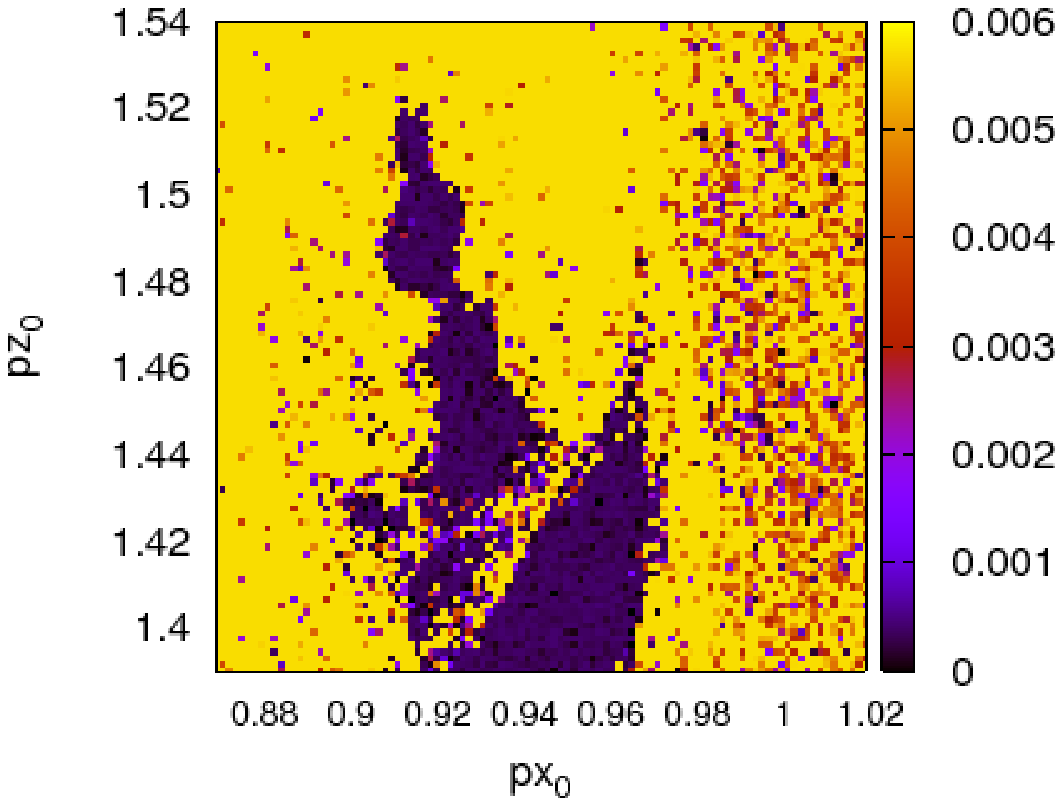}}\\
\resizebox{75mm}{!}{\includegraphics{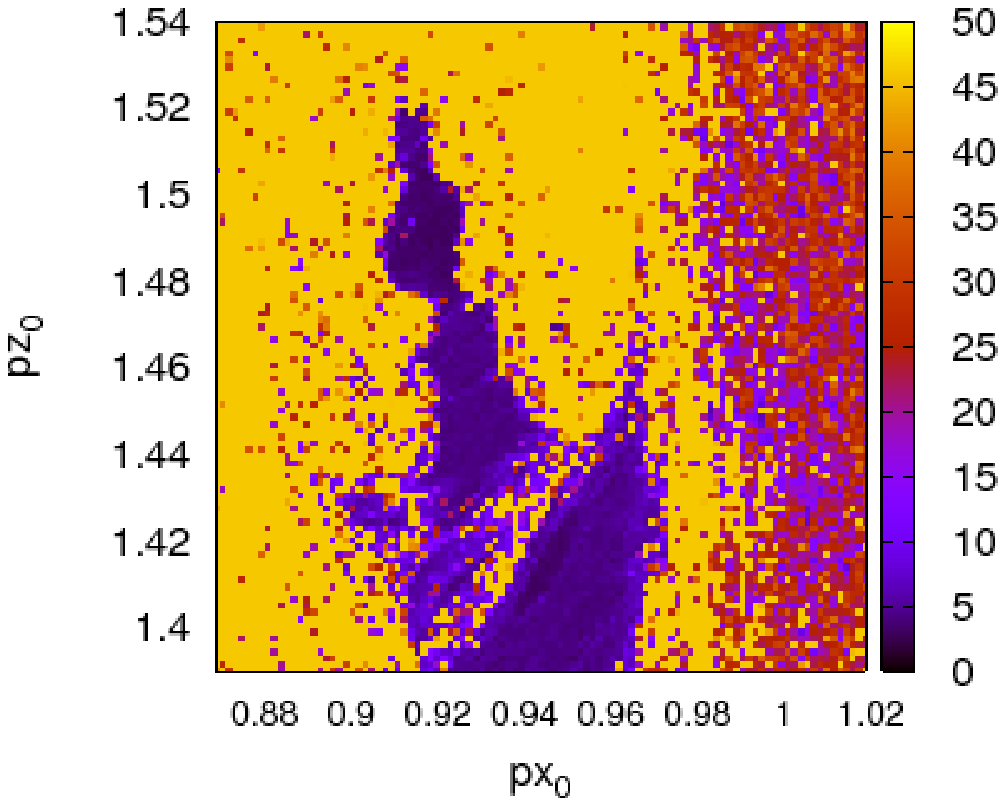}}&
\resizebox{75mm}{!}{\includegraphics{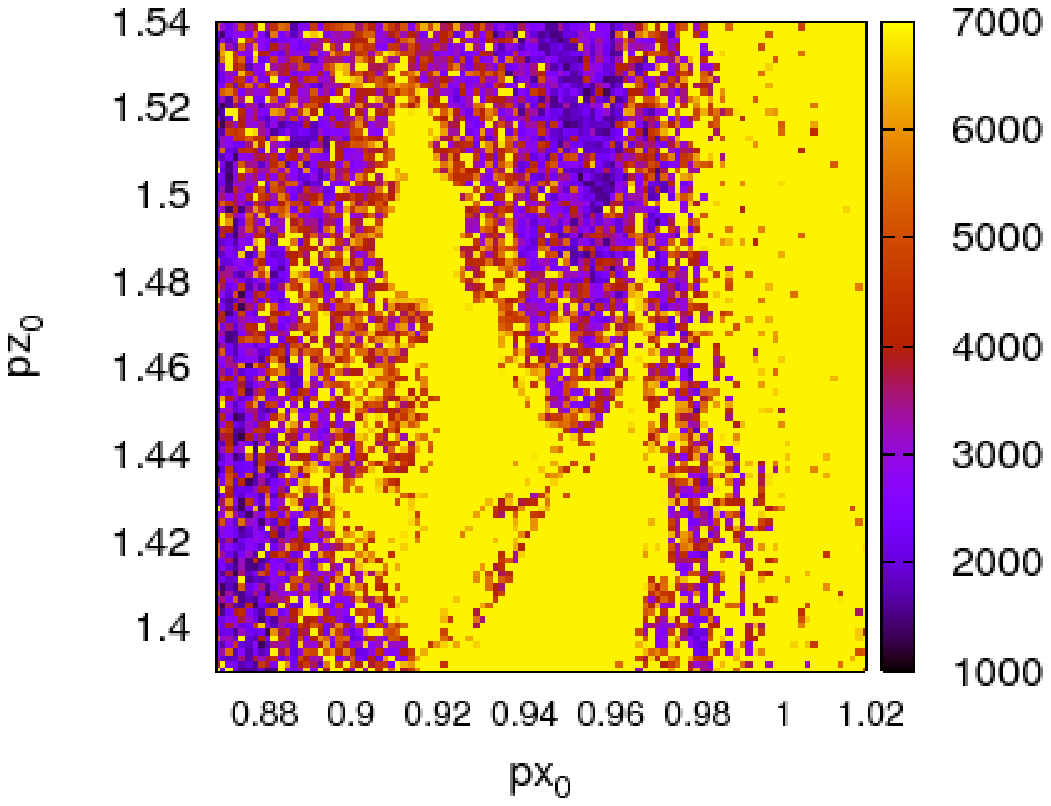}}\\
\resizebox{75mm}{!}{\includegraphics{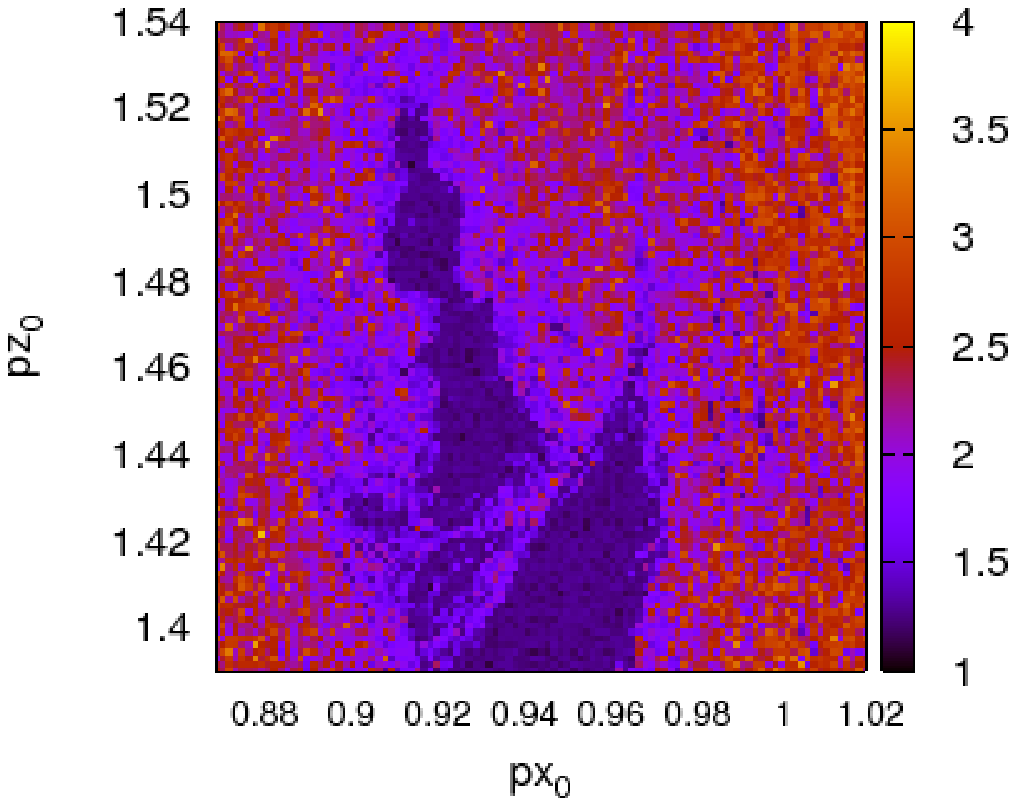}}&
\resizebox{75mm}{!}{\includegraphics{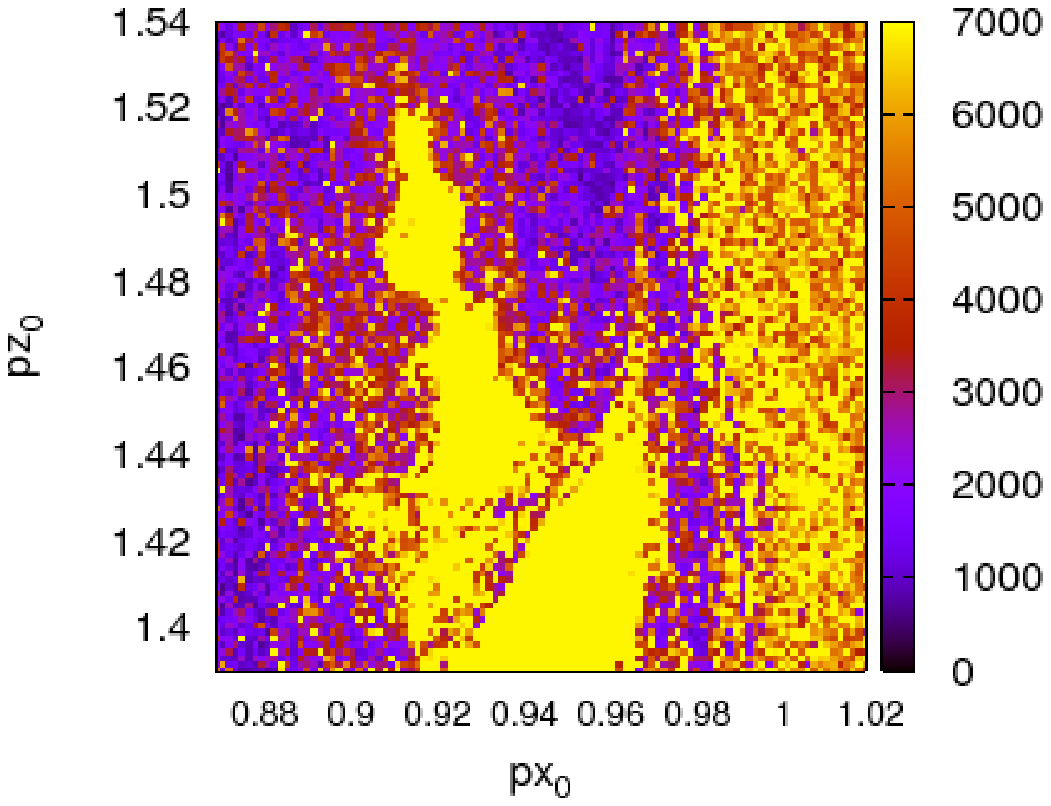}}
\end{tabular}
\caption{We show the phase space portraits of the different VICs using a total integration time of $7\times 10^3$ u.t. for the region displayed on Fig. \ref{region-in-stationary}. We present the final values of the MEGNO (top left panel); the final values of the SE\textit{l}LCE (top right panel); the final values (middle left panel, in logarithmic scale) and the saturation times (middle right panel) of the OFLI; and the final values (bottom left panel) and the saturation times (bottom right panel) of the APLE. The cold colours correspond to regular regions while the warm colours correspond to chaotic domains for the final values. The colour scale is inverted for the saturation times.} 
\label{STATIONARY-1}
\end{figure*}

In Fig. \ref{STATIONARY-1} we present the phase space portraits given by the final values of the MEGNO and the SE\textit{l}LCE (top left and top right panels, respectively); the OFLI final values (middle left panel) and the saturation times (middle right panel); and the APLE final values (bottom left panel) and the saturation times (bottom right panel). The phase space portraits given by the FLI are very similar to those shown with the OFLI and they are not included for the sake of simplicity.

The region under analysis has a regular component (the main central structure) and a chaotic component (surrounding the main central structure). Moreover, there are two different regions of chaos. The region that surrounds the regular component is composed of very strong chaotic orbits while the band on the right side of every panel in Fig. \ref{STATIONARY-1} is filled with moderate to strong chaotic orbits.  

The values of the MEGNO and the SE\textit{l}LCE do not grow exponentially for chaotic orbits as it happens, for instance, with the OFLI (and the FLI) or the APLE. Then, a saturation value is not defined and a saturation time is not recorded in the computing process. However, the values of the MEGNO and the SE\textit{l}LCE can be very large for strong chaotic orbits and very different from their values for regular and mild chaotic orbits. In a divided phase space with regions of very strong chaos, the large values of these indicators for strong chaotic orbits may hide the small differences between regular and mild chaotic orbits. Indeed, the differences among strong chaotic orbits are generally not as important as the differences between strong and mild chaotic orbits or between mild chaotic orbits and regular orbits. Thus, a saturation value for the MEGNO or the SE\textit{l}LCE is also very useful in this kind of scenario. As shown in Fig. \ref{region-in-stationary}, we take the saturation value $20$ for the MEGNO and its associated value $\sim0.0057$ for the SE\textit{l}LCE\footnote{We can estimate the saturation value ($\tilde{\chi}_{\gamma}$) for the SE\textit{l}LCE from the expression: $20\sim\tilde{\chi}_{\gamma}/2\times t$, with $t=7\times 10^3$ u.t. (see Section \ref{CIs&FMFT-S1}).}. 

In case of the OFLI (and the FLI) or the APLE, the chaotic orbits grow exponentially fast and a saturation value is needed to avoid the propagation of errors in their computation. The saturation value in Fig. \ref{STATIONARY-1} for the OFLI (middle panels) is $\sim\ln(10^{20})$ (see Section \ref{CIs&FMFT-S2}). The saturation value in Fig. \ref{STATIONARY-1} for the APLE (bottom panels) is $\sim\ln(10^{20})/\ln(t)$ (see Section \ref{CIs&FMFT-S3}).

In the top panels of Fig. \ref{STATIONARY-1}, we can see the regular and the chaotic components well separated by the MEGNO (left panel) and the SE\textit{l}LCE (right panel). Furthermore, the two different chaotic regions are clearly shown: the region of strong chaos (surrounding the central structure with regular orbits) and the band on the right side of the panels filled with moderate to strong chaotic orbits. There is no structure at all in the main central structure composed of regular orbits (top left panel) because of the MEGNO's asymptotic threshold (Section \ref{CIs&FMFT-S1}). Also, there is no further structure in the strong chaotic region due to the selected saturation value. The SE\textit{l}LCE shows a similar portrait (top right panel).    

In middle panels of Fig. \ref{STATIONARY-1}, we use the OFLI to describe the region under analysis. On the one hand, we recover the phase space portraits (middle left panel) given by the MEGNO and the SE\textit{l}LCE in the top panels of the same figure. On the other hand, the band on the right side of both panels shows a more detailed structure with the OFLI. This is due to the fact that the saturation value taken for the OFLI covers a wider range of chaotic orbits than the values chosen for the MEGNO or the SE\textit{l}LCE. The saturation times (middle right panel) enhance the information given with the final values of the OFLI, specially within the strong chaotic region. 

In the bottom panels of Fig. \ref{STATIONARY-1}, we present the phase space portraits given by the APLE final values (left panel) and saturation times (right panel). The description is very similar to those given by the other VICs if we consider both the final values and the saturation times of this indicator. For example, the identification of the band on the right side of the panels needs the information of the saturation times. The saturation times show a clear distinction between the region of strong chaos surrounding the main central structure and the region with mild and strong chaotic orbits in the band on the right side of the panels. This is not the case for the other VICs (the MEGNO, the SE\textit{l}LCE, the OFLI and the FLI) for which the saturation times play a complementary role and the final values were enough for such identification. Nevertheless, the degree of coincidence in the descriptions of the four VICs (plus the FLI) seems to be high. Thus, there is no decisive advantage in favour of one of them.  

\subsection{The speed of convergence of the VICs: thresholds}\label{STATIONARYS2} 
Now we focus on another important characteristic of the VICs which is fundamental for an efficient study of large samples of i.c.: the speed of convergence. 

We use the same total integration time ($7\times 10^3$ u.t.) and the same i.c. (10201) used in Section \ref{STATIONARYS1}. The initial $10^2$ u.t. of the time interval does not give reliable information for our purposes and it is not considered in this experiment. The remaining $6.9\times 10^3$ u.t. are divided into $70$ sub-intervals of $\sim98.57$ u.t. each. We compute the number of chaotic orbits in each sub-interval. Before that, we had to identify the chaotic orbits. So, we needed accurate thresholds for each VIC. Many of them have theoretical or empirical estimations of their thresholds. Yet, they are only approximations and should be adjusted to each particular situation. Hence, we generalize the experiments shown in M11 and D12 and consider a wide range of thresholds for the different VICs. Finally, for each sub-interval, we compute the number of chaotic orbits for every VIC according to the different thresholds. 

Performing several tests (i.e. considering different thresholds and VICs), we find that a percentage of $\sim83\%$ for the chaotic component is stable and prevails in the experiment. We take it as the ``true'' percentage for the chaotic component. Having such a ``true'' percentage, we can adjust the thresholds of the VICs under consideration. The fastest convergency of the VIC to a stable percentage of chaotic orbits close to the ``true'' percentage gives the most appropriate threshold for that particular indicator.  

\begin{figure*}
\begin{tabular}{cc}
\resizebox{75mm}{!}{\includegraphics{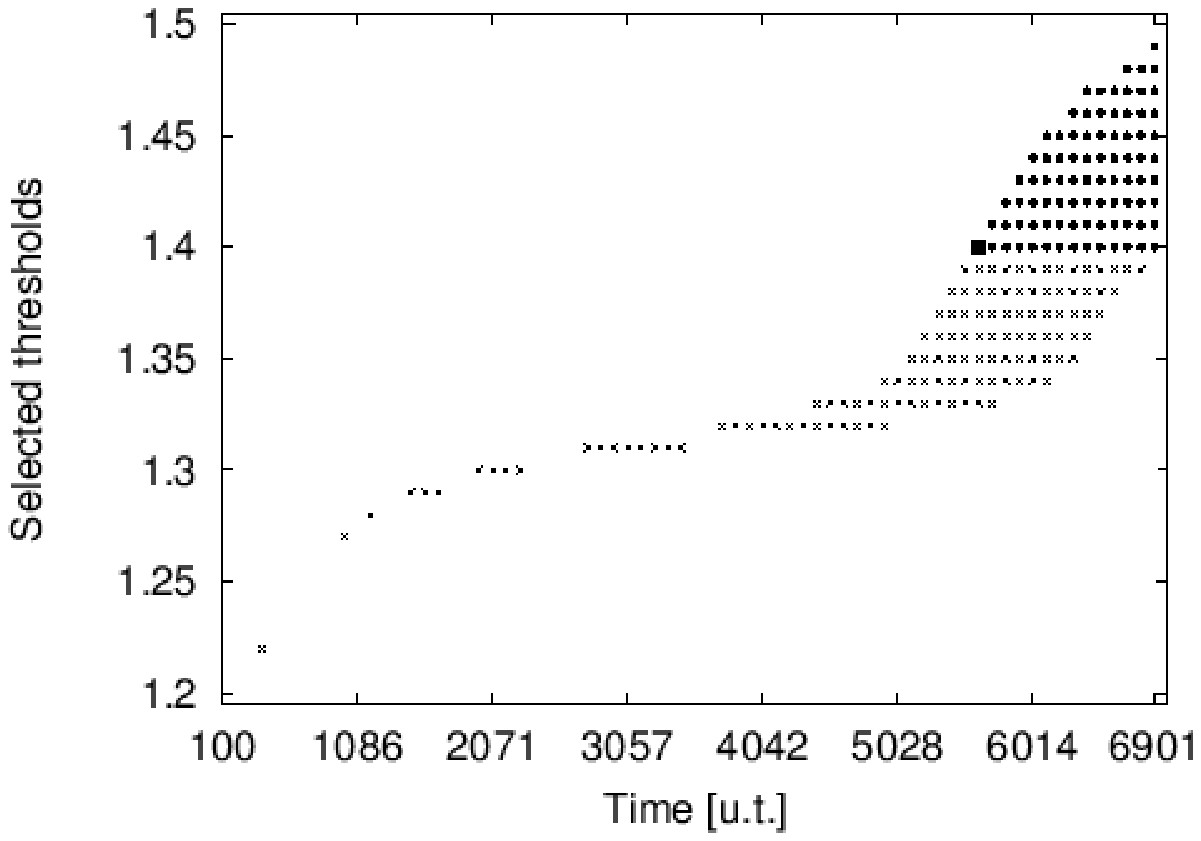}}&
\resizebox{75mm}{!}{\includegraphics{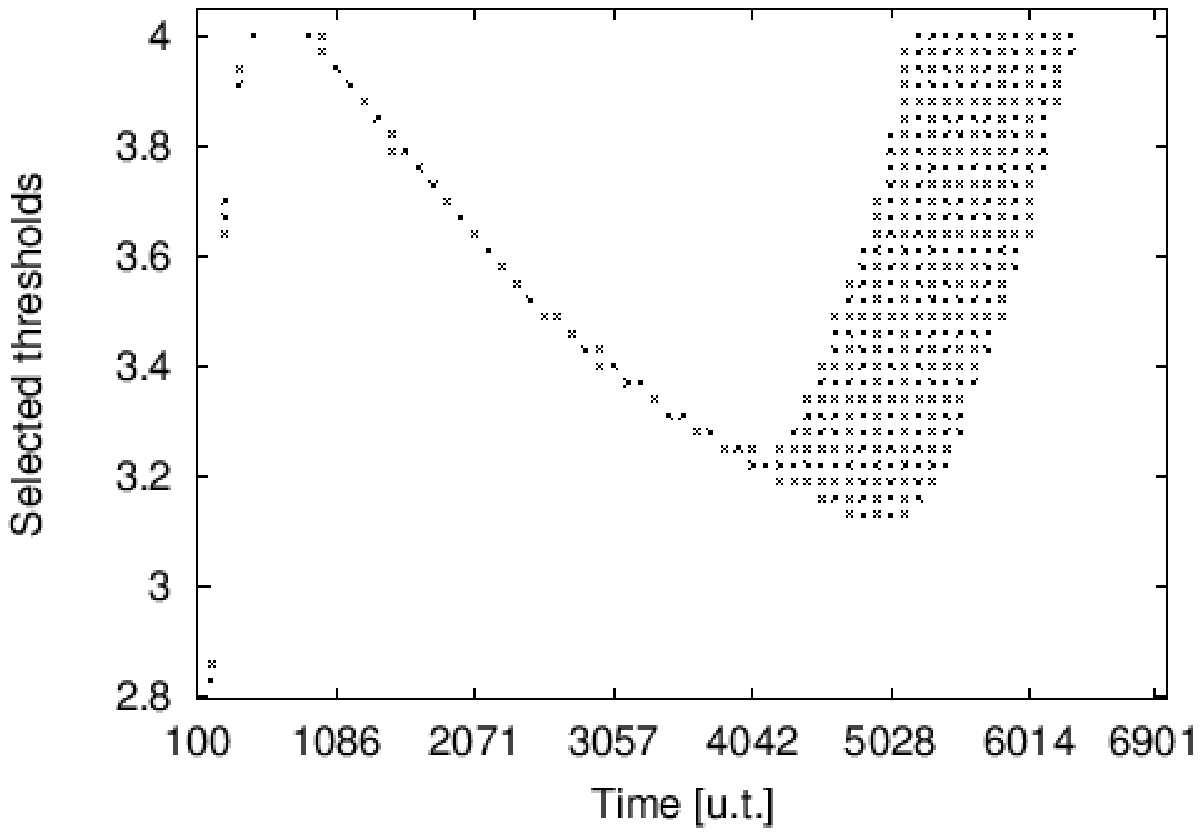}}\\ 
\resizebox{75mm}{!}{\includegraphics{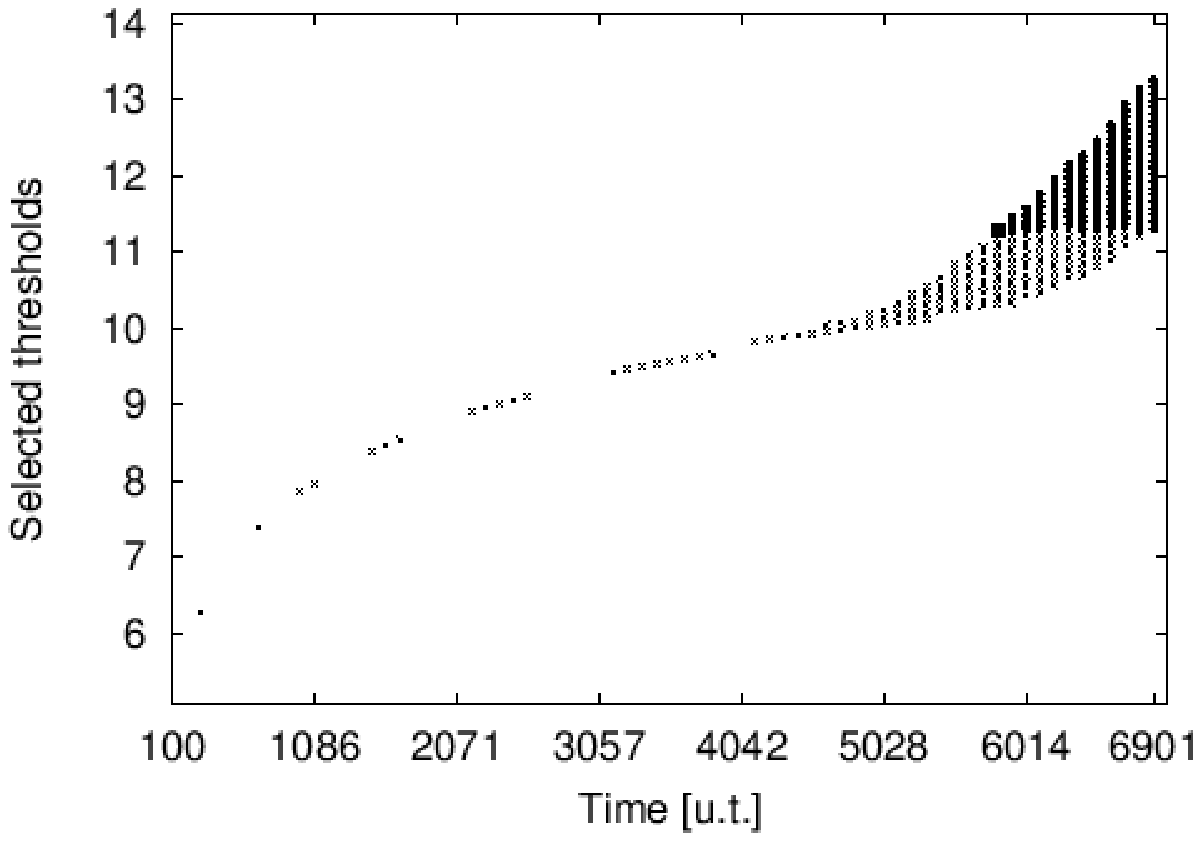}}&
\resizebox{75mm}{!}{\includegraphics{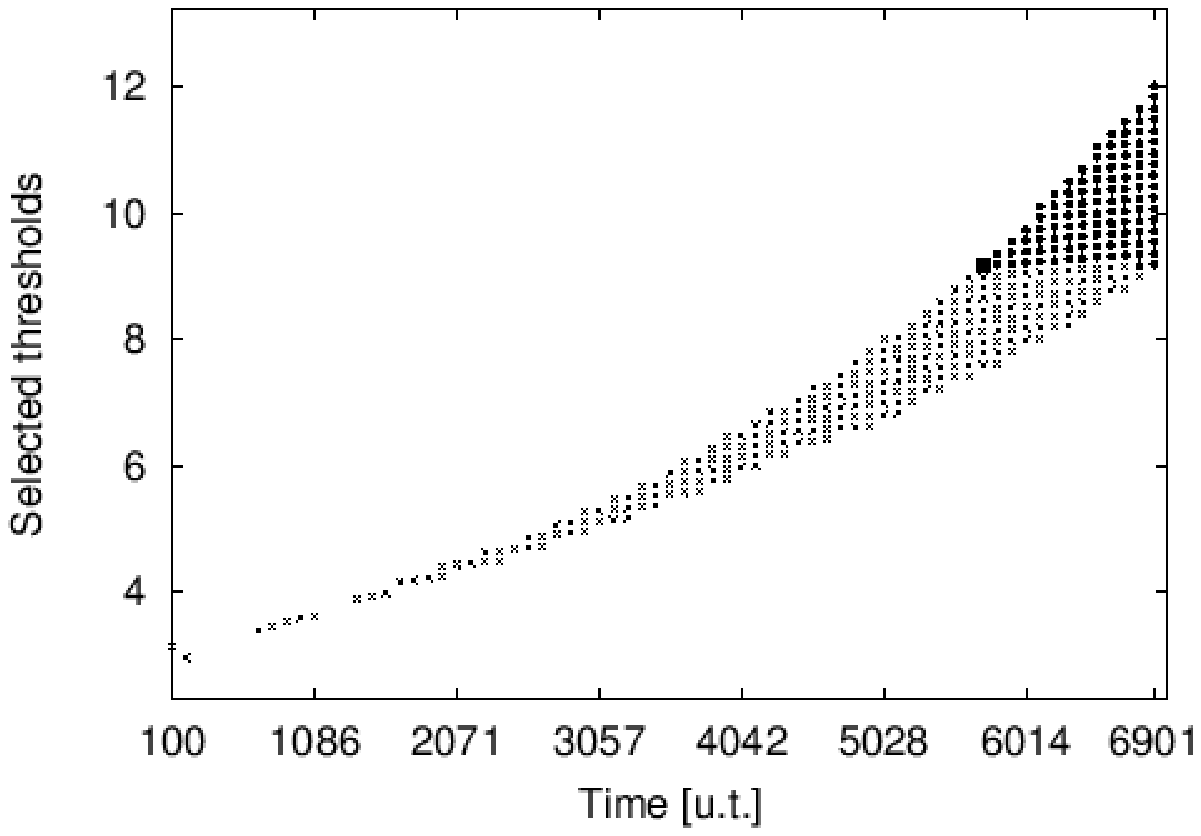}}
\end{tabular}
\caption{We present the selected thresholds for the APLE (top left panel), the MEGNO (top right panel), the FLI (bottom left panel) and the OFLI (bottom right panel). The sample consists of 10201 i.c. for $7\times 10^3$ u.t. in the region shown in Fig. \ref{region-in-stationary}. See text for further details.}
\label{STATIONARY-2}
\end{figure*}

In order to identify such appropriate thresholds for every VIC under consideration, we define the ``selected thresholds''. These thresholds allow the associated indicator to reach the ``true'' percentage of chaotic orbits within the total integration time\footnote{Actually, the selected thresholds allow the associated VIC to reach a percentage of chaotic orbits between 82-84\% within the total integration time.}. 

In Fig. \ref{STATIONARY-2} we present the variation of the selected thresholds with the integration time for the APLE (top left panel), the MEGNO (top right panel), the FLI (bottom left panel) and the OFLI (bottom right panel). The SE\textit{l}LCE is not included in the experiment because it has not a defined mechanism to compute its threshold. The filled circles in Fig. \ref{STATIONARY-2} indicate the selected thresholds that allow the convergency of the corresponding VIC to a stable, close to the ``true'' percentage of chaotic orbits. The filled squares in the same figure indicate the selected thresholds that offer the fastest convergency. In other words, the filled squares indicate the most appropriate thresholds for a given indicator. 

The APLE has the theoretical time--independent threshold $\sim1$ (see Section \ref{CIs&FMFT-S3} and \citet{LVE08} for further details). However, we consider threshold values between $\sim1.2$ and $\sim1.5$. The threshold values below $\sim1.2$ characterise 100\% of the orbits as chaotic orbits for the whole time interval and thus, they are discarded. The threshold values above $\sim1.5$ do not arrive at the ``true'' percentage of the chaotic component within the time interval and thus, they are also discarded. If we use the threshold values in the interval $\sim[1.4:1.5]$, the APLE converges to stable percentages of the chaotic component close to the ``true'' percentage (see top left panel of Fig. \ref{STATIONARY-2}). If we use the threshold values around $1.4$, the APLE's convergency is the fastest (approximately by $5.5\times 10^3$ u.t.).

The MEGNO, like the APLE, has also a time--independent threshold: $\sim2$ (see Section \ref{CIs&FMFT-S1} and \citet{CGS03} for further details). For the same reasons given for the APLE, we consider threshold values above the theoretical approximation, i.e. between $\sim2.8$ and $\sim4$. The threshold values below $\sim2.8$ characterise 100\% of the orbits as chaotic orbits for the whole time interval and thus, they are discarded. The threshold values above $\sim4$ are much larger than the theoretical asymptotic estimation of $\sim2$ and thus, they are also discarded. If we use the threshold values close to $\sim4$, the MEGNO reaches percentages of the chaotic component close to the ``true'' percentage more rapidly (see top right panel of Fig. \ref{STATIONARY-2}). Nevertheless, using this range of selected thresholds the percentages are not convergent. Moreover, the percentages oscillate within the interval $\sim[84\%:87\%]$ depending on the threshold selected. In other words, the MEGNO gives a higher percentage of chaotic orbits than the prevailing percentage shown by the other VICs. This experiment confirms the results yielded in M11 and D12. 

The FLI and the OFLI have both similar behaviour and time--dependent thresholds to distinguish chaotic motion from regular motion (see Section \ref{CIs&FMFT-S2} and \citet{FLFF02} for further details). After computing the logarithm in their definitions, a tentative threshold for them is $\ln(t)$ with $t$ the integration time. We consider the general expression $a\times\ln(t)$ in order to study a variation of the thresholds changing the values of the parameter $a$. 

The values of the parameter $a$ for the FLI belong to the interval $\sim[1.2:1.6]$. The theoretical threshold with $a\sim1$ is not efficient because it does not arrive at stable percentages of the chaotic component close to the ``true'' percentage for the whole time interval. The same applies to the thresholds with $a\lesssim1.2$ so they are discarded. Furthermore, the threshold values with $a\gtrsim1.6$ do not reach at the ``true'' percentage of the chaotic component within the time interval and thus, they are also discarded. If we use the threshold values in the interval $\sim[1.2:1.6]$ in the time interval $\sim[5.7\times10^3:7\times10^3]$ u.t., the FLI converges to stable percentages of the chaotic component close to the ``true'' percentage (see the bottom left panel of Fig. \ref{STATIONARY-2}). If we use the threshold values with $a\sim1.3$, the FLI's convergency is the fastest (approximately by $5.7\times 10^3$ u.t., i.e. $1.3\times\ln(5.7\times 10^3)\sim11.2$ in the figure). These results are very similar to those obtained earlier with the APLE. Yet, there is an important difference: the threshold selected for the FLI ($a\sim1.3$) is closer to the theoretical approximation ($a\sim1$) than the threshold selected for the APLE ($a\sim1.4$). Therefore, this indicates that the FLI may have a better theoretical approximation of its threshold than the APLE.

Finally, we also observe similar results for the OFLI. However, the parameter $a$ varies from $\sim0.5$ to $\sim1.5$, i.e., for the first time the interval of selected thresholds includes the theoretical estimation ($a\sim1$). That is, if we use the theoretical approximation, the OFLI reaches a stable percentage of chaotic orbits close to the ``true'' percentage of $83\%$. Nevertheless, the convergency arrives at the end of the time interval, i.e. $\sim6.9\times10^3$ u.t. The threshold values below $\sim0.5$ characterise 100\% of the orbits as chaotic orbits for the whole time interval and thus, they are discarded. The threshold values above $\sim1.5$ do not arrive at the ``true'' percentage of the chaotic component within the time interval and thus, they are also discarded. If we use the thresholds with $a$ in the interval $\sim[1:1.4]$ and within the time interval $\sim[5.7\times10^3:7\times10^3]$ u.t., the OFLI converges to stable percentages of the chaotic component close to the ``true'' percentage (see the bottom right panel of Fig. \ref{STATIONARY-2}). If we use the threshold values with $a\sim1.07$, the OFLI's convergency is the fastest (approximately by $5.7\times 10^3$ u.t., i.e. $1.07\times\ln(5.7\times 10^3)\sim9.2$ in the figure). These results are very similar to those obtained with the FLI as mentioned before. Yet, once again there is an important difference: the threshold selected for the OFLI ($a\sim1.07$) is closer to the theoretical approximation ($a\sim1$) than the threshold selected for the FLI ($a\sim1.3$). Therefore, this indicates that the theoretical approximation of the thresholds of the OFLI may work better.

\section{Global analysis of the start spaces}\label{thespaces}
In this section we proceed to study some dynamical aspects of the ScTS model by means of the complementary use of the VICs selected in Section \ref{STATIONARY}. 

We consider samples of around one millon i.c. on the $p_{x_0}-p_{z_0}$ and the $x_0-z_0$ start spaces for four different energy surfaces. Even though we do not study the remaining five start spaces indicated in \citet{S93,PL98}, we gain sufficient information in order to briefly discuss some main features of the dynamics of the system.

The total integration times extend to $1.17\times 10^5$ u.t. for the energy surface $-0.1$ (Section \ref{FMFT-CIS1}). The CPU times become critical and will be considered a priority in the selection of the VICs. The FLI family of indicators presents the most versatile indicators, showing good performances in all the experiments (see also M11 and D12 in which they are included in the corresponding CIsF). The MEGNO is also another candidate because of its good performance (also included in the CIsF of M11 and D12). The SE\textit{l}LCE seems to be a good alternative to the MEGNO when studying big samples of orbits. Furthermore, the MEGNO and the SE\textit{l}LCE show a lower computational cost in the experiment than that of the FLI family of indicators\footnote{In this test on computing times, we consider the definition of the FLI given in \citet{FL00}, where the authors compute the evolution of the FLI using one deviation vector.}. Therefore, we select the LI (the least time--consuming indicator given a fixed total integration time, see e.g. D12) and the pair MEGNO/SE\textit{l}LCE as the VICs to study the ScTS model.

\subsection{The $p_{x_0}-p_{z_0}$ start space}\label{STATIONARYS4} 
\begin{table*}
\centering
\caption{For every energy surface on the $p_{x_0}-p_{z_0}$ start space, we present the threshold associated to the LI and the estimated percentage of chaotic orbits; the threshold associated to the SE\textit{l}LCE and the estimated percentage of chaotic orbits.}
\begin{tabular}{@{}ccccc@{}}
\hline
Energy surface & Threshold (LI) & Chaotic orbits (\%) -LI &  Threshold (SE\textit{l}LCE) & Chaotic orbits (\%) -SE\textit{l}LCE\\ 
\hline 
 $-0.1$ & $1.2\times10^{-4}$ & $\sim71.46\%$ & $1.4\times10^{-5}$ & $\sim66.83\%$ \\
 $-0.3$ & $5.8\times10^{-4}$ & $\sim65.89\%$ & $6.4\times10^{-5}$ & $\sim65.99\%$ \\
 $-0.5$ & $1.25\times10^{-3}$ & $\sim61.79\%$ & $1.3\times10^{-4}$ & $\sim63.05\%$\\
 $-0.7$ & $1.7\times10^{-3}$ & $\sim65.09\%$ & $2.7\times10^{-4}$ & $\sim62.87\%$ \\
\hline
\end{tabular}
\label{tablestationary-percent}
\end{table*}

\begin{figure*}
\begin{tabular}{cc}
\resizebox{75mm}{!}{\includegraphics{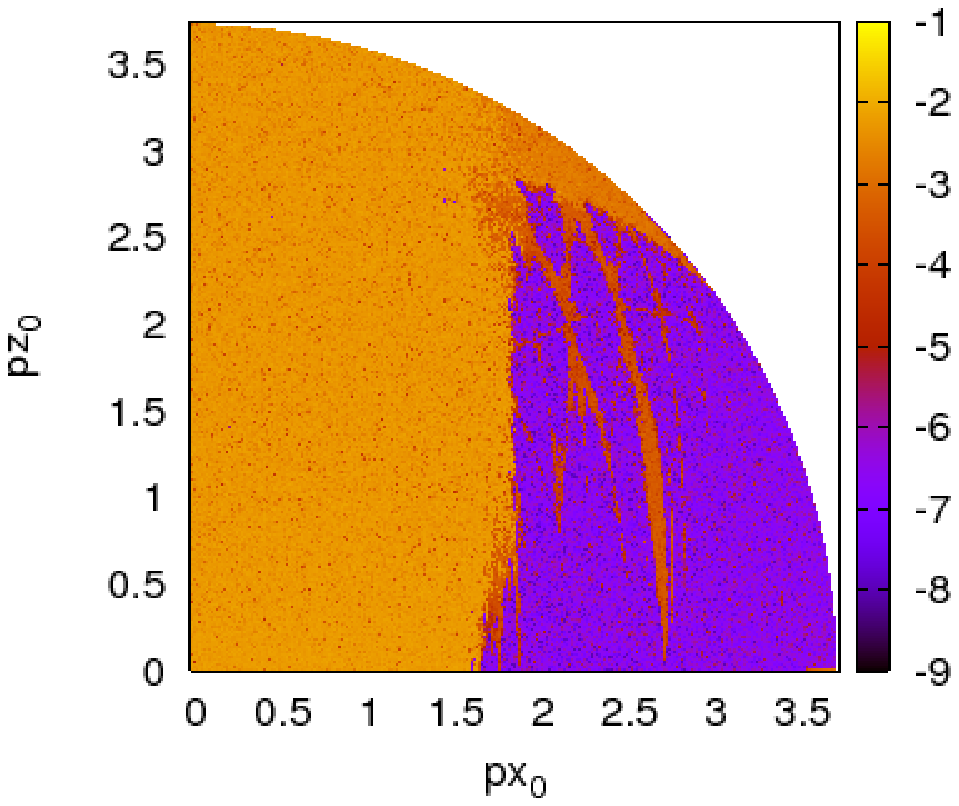}}& 
\resizebox{75mm}{!}{\includegraphics{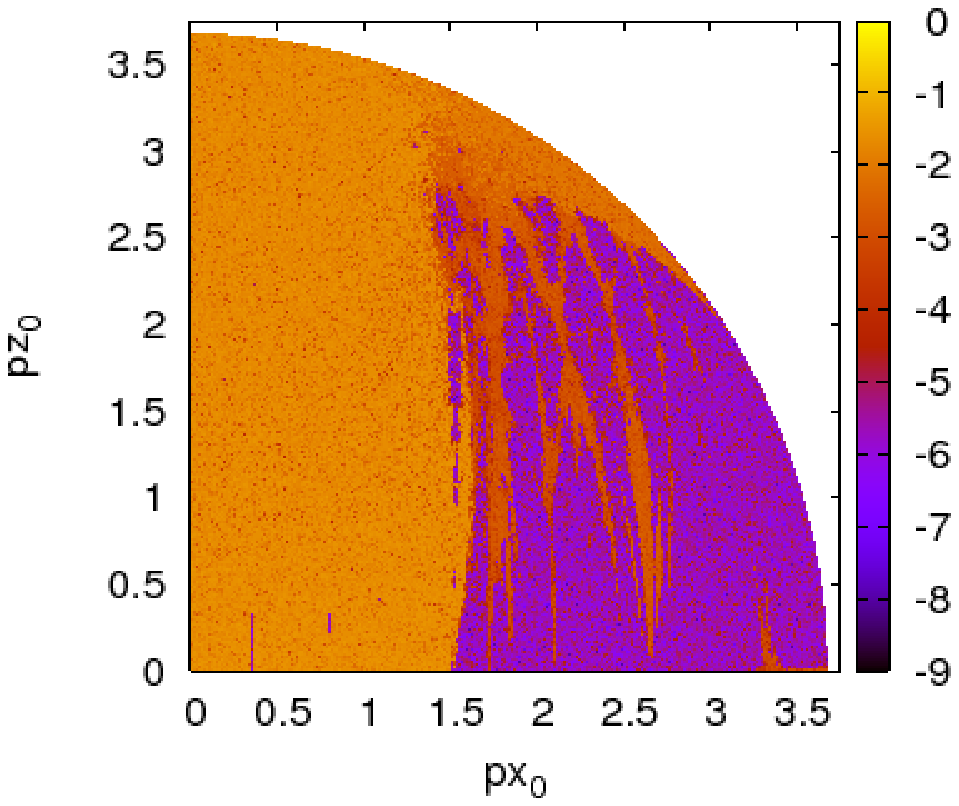}}\\
\resizebox{75mm}{!}{\includegraphics{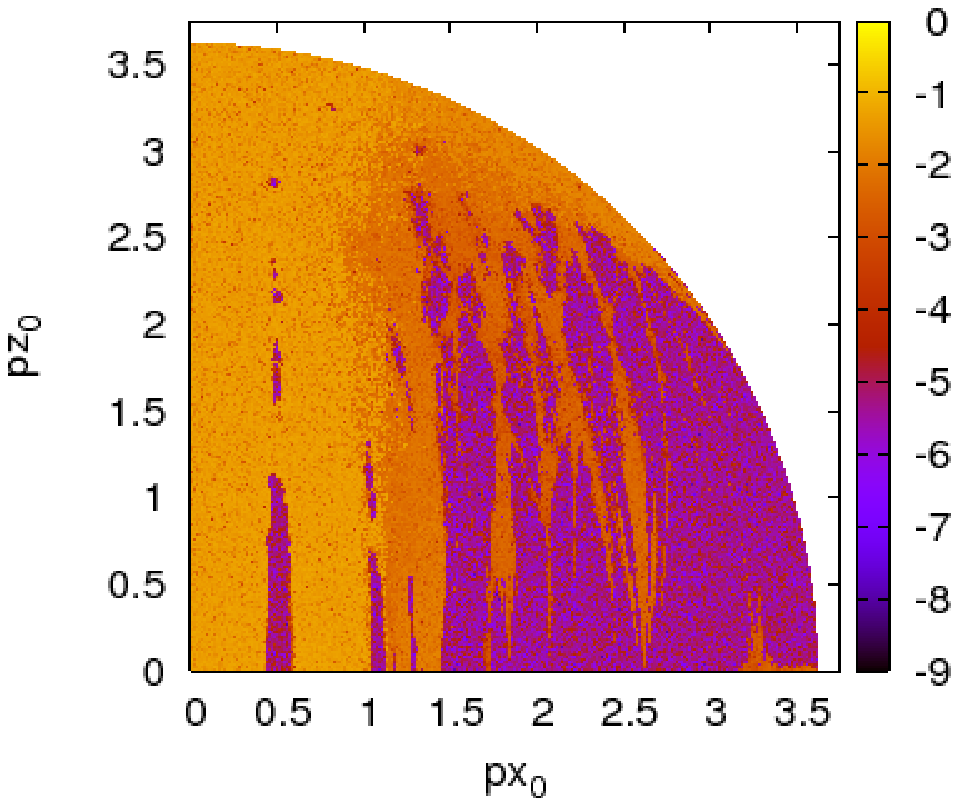}}& 
\resizebox{75mm}{!}{\includegraphics{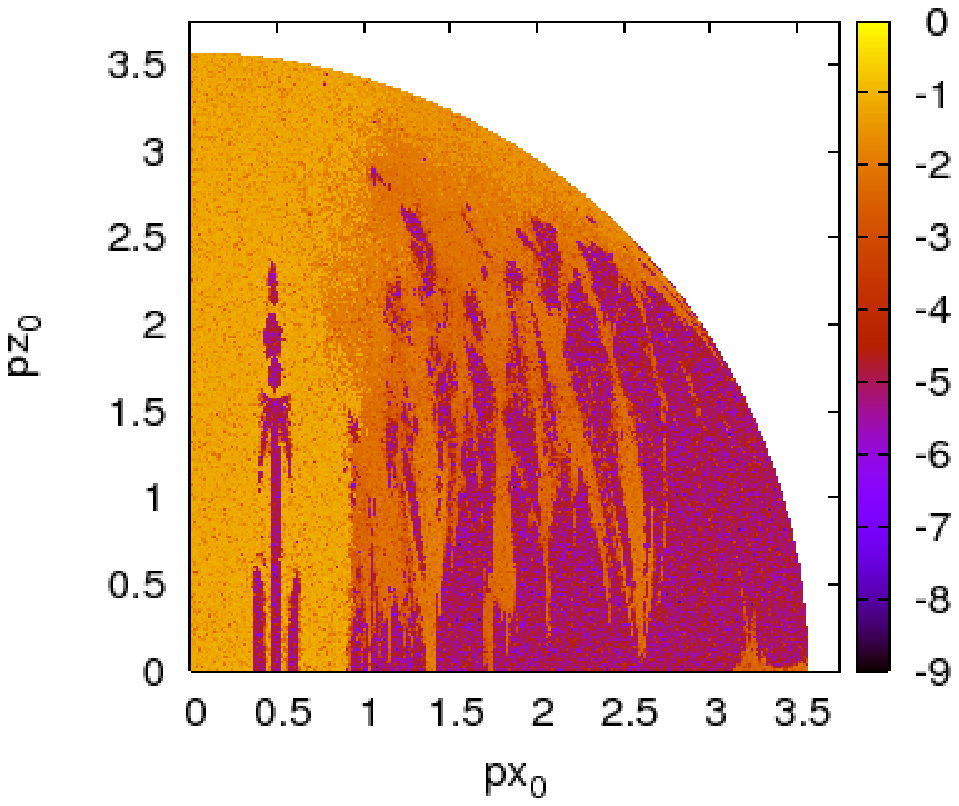}}
\end{tabular}
\caption{Phase space portraits of the SE\textit{l}LCE for a sample of 1000444 i.c. on the $p_{x_0}-p_{z_0}$ start space and for four energy surfaces. Top left panel: phase space portrait corresponding to the energy surface $-0.1$ and using a total integration time of $1.17\times 10^5$ u.t. Top right panel: $-0.3$ and $2.4\times 10^4$ u.t. Bottom left panel: $-0.5$ and $1.2\times 10^4$ u.t. Bottom right panel: $-0.7$ and $7\times 10^3$ u.t. Chaotic regions are identified with values of the indicator higher than $\sim-5$ or $\sim-4$ for the energy surfaces $-0.1$ and $-0.3$ or $-0.5$ and $-0.7$, respectively. See Table \ref{tablestationary-percent} for further details. Notice that the values of the indicator are in logarithmic scale.}
\label{C5S4S1-4}
\end{figure*}

We have already mentioned that the SE\textit{l}LCE has certain advantages when we study big samples of orbits (see Section \ref{CIs&FMFT-S1}). Nevertheless, it does not have any defined procedure to determine a threshold in order to distinguish between regular and chaotic motion. Indeed, this fact is a serious problem if we want to separate regular motion from weak chaotic motion. We need a threshold for the SE\textit{l}LCE to correctly describe the $p_{x_0}-p_{z_0}$ start space of the ScTS model through the final values of the indicator. For the determination of such a threshold we require again a ``true'' percentage of chaotic orbits for the sample under study. The MEGNO shows high sensitivity to changes in its threshold (Section \ref{STATIONARYS2}, M11 and D12) and hence it is not a reliable indicator to calibrate other VICs. Besides, the LI has a theoretical threshold ($\ln(t)/t$, with $t$ the total integration time), which is a still valid estimation as a decent first approximation. We start with the corresponding theoretical estimations\footnote{The theoretical estimations: $9.9743\times10^{-5}$ (u.t.)$^{-1}$ for the energy surface $-0.1$; $4.20242\times10^{-4}$ (u.t.)$^{-1}$ for $-0.3$; $7.82722\times10^{-4}$ (u.t.)$^{-1}$ for $-0.5$ and $1.264809\times10^{-3}$ (u.t.)$^{-1}$ for $-0.7$ give overestimated percentages of the chaotic component.} and calibrate them using inspections of different profiles of i.c. within the sample. Then, we find reliable thresholds for the LI for every energy surface. We compute the ``true'' percentages of chaotic orbits with the LI and calibrate the thresholds of the SE\textit{l}LCE to best fit these values. In Table \ref{tablestationary-percent}, we present the results ordered by energy surface. Notice that the thresholds associated with the SE\textit{l}LCE are considerably smaller than the thresholds of the LI, because the estimation of the lLCE given by the slope of the MEGNO converges more rapidly to the lLCE than the standard procedure (see Section \ref{CIs&FMFT-S1} and \citet{CGS03}). 

A comment regarding the efficient way to work with the LI and the SE\textit{l}LCE is in order before going forward. On one hand the LI is a reliable, but slow indicator so that it should be applied for longer integration times but on smaller samples in order the computing times to be moderate. Then, we can estimate the percentages of chaotic orbits. On the other hand, we have the SE\textit{l}LCE, which has a fast speed of convergence so it can be applied for shorter integration times but on the whole sample of orbits (once again, the computing times are reasonably short). Finally, we calibrate the thresholds of the SE\textit{l}LCE adjusting the percentages of chaotic orbits to those obtained with the LI. This is an efficient way to take advantage of the different characteristics (reliability and speed of convergence) of both VICs. Nevertheless, the priority in this experiment is to obtain reliable percentages of chaotic orbits rather than saving computing time in order to obtain the more accurate phase space portraits. This is the reason why we previously applied both the LI and the SE\textit{l}LCE on the same samples and for the same total integration times.

Now that we have the thresholds for the SE\textit{l}LCE, we apply the latter to a sample of 1000444 i.c. to study the $p_{x_0}-p_{z_0}$ start space for the following four energy surfaces\footnote{We use a cut--off level around $10^{-9}$ to have the same scales in Fig. \ref{C5S4S1-4} for every energy surface. Then, the percentages of orbits that we do not consider are negligible. Those percentages are $\sim0.77$\% in the energy surface $-0.1$, $\sim0.06$\% in $-0.3$, $\sim0.03$\% in $-0.5$ and $\sim0.02$\% in $-0.7$.}: $-0.1$, $-0.3$, $-0.5$ and $-0.7$ (see Fig. \ref{C5S4S1-4}).

In Fig. \ref{C5S4S1-4} we observe that independently of the energy surface, the chaotic component is the dominant one on the phase space portraits (yellow and orange colours. See also column three and five in Table \ref{tablestationary-percent}). The regular component (red, blue and black colours) increases for lower energy values. However, this increment is not considerable yet. 

On the top left panel of Fig. \ref{C5S4S1-4}, we observe a clear separation of both components. We show a fully connected chaotic component for values of $p_{x_0}\lesssim1.7$ and the regular component for values of $p_{x_0}\gtrsim1.7$, besides some hyperbolic structures. 

These structures grow as we move to lower energy surfaces. They are resonances that start to overlap with each other and occupy the regular component. We can also distinguish a separation inside the chaotic component. The fully connected chaotic domain moves back to lower values of $p_{x_0}$. The region is occupied by another chaotic domain characterised by a resonance overlapping regime and a lower Lyapunov number (top right panel of Fig. \ref{C5S4S1-4}). On the energy surface $-0.5$ (bottom left panel of Fig. \ref{C5S4S1-4}), the fully connected chaotic domain has noticeably moved backwards and the resonances occupy extended regions of the $p_{x_0}-p_{z_0}$ start space. Furthermore, a major resonance is seen around the value $p_{x_0}\sim0.5$ and minor resonances are seen around the values $p_{x_0}\sim1$ and $p_{x_0}\sim1.3$. On the bottom right panel of Fig. \ref{C5S4S1-4}, for the energy surface $-0.7$, the portrait shows how the resonances occupy most of the regular component. Moreover, major resonances grow within the chaotic domains. The most evident resonance is that located at value $p_{x_0}\sim0.5$, which has already been identified in outer energy surfaces. 

In spite of the short discussion of the $p_{x_0}-p_{z_0}$ start space given above, the efficiency of variational indicators to describe the global characteristics of the phase space is evident. The VICs easily identify the regular and the chaotic domains and their mutual interplay, along with phase space mixing processes like resonance overlapping. However, the identification of the underlying resonant structure by the frequency vectors is not readily understood. We should study the dimensionality of the tori (e.g. with the GALI method, \citet{SBA07}, D12) to identify resonant orbits, locate the periodic orbits (e.g. with the OFLI, \citet{FLFF02}, D12) and analyse their stability (e.g. with the MEGNO, \citet{CGM08}) to look for resonant orbit families. The process could be time--consuming and rather inefficient since the VICs are not the most suitable tools for the task. In fact, the SAMs are far better options (see Section \ref{intro} and references therein).

The dynamics of the $p_{x_0}-p_{z_0}$ start space can be briefly described as a dominant chaotic component based mostly on a resonance overlapping regime. This phase mixing process produces a more homogeneous chaotic component for the energy surface $-0.1$.  

For further studies concerning the $p_{x_0}-p_{z_0}$ start space of the ScTS model refer to \citet{MCW05} and \citet{CGM08}.

Now, we apply the same package of techniques to the $x_0-z_0$ start space, where the dynamics is totally different since it is dominated by the regular orbits. 

\subsection{The $x_0-z_0$ start space}\label{STARTX0Z0}
\begin{table*}
\centering
\caption{For every energy surface in the $x_0-z_0$ start space, we present the threshold associated to the LI and the estimated percentage of chaotic orbits; the threshold associated to the SE\textit{l}LCE and the estimated percentage of chaotic orbits.}
\begin{tabular}{@{}ccccc@{}}
\hline
Energy surface & Threshold (LI) & Chaotic orbits (\%) -LI &  Threshold (SE\textit{l}LCE) & Chaotic orbits (\%) -SE\textit{l}LCE\\ 
\hline 
 $-0.1$ & $1.25\times10^{-4}$ & $\sim3.76\%$ & $9.4\times10^{-6}$ & $\sim4.65\%$ \\
 $-0.3$ & $5.3\times10^{-4}$ & $\sim8.15\%$ & $3.2\times10^{-5}$ & $\sim9.2\%$ \\
 $-0.5$ & $9.4\times10^{-4}$ & $\sim15.06\%$ & $6.5\times10^{-5}$ & $\sim12.74\%$ \\
 $-0.7$ & $1.7\times10^{-3}$ & $\sim16.37\%$ & $9.5\times10^{-5}$ & $\sim16.75\%$ \\
\hline
\end{tabular}
\label{tablestart-percent}
\end{table*}

\begin{figure*}
\begin{tabular}{cc}
\resizebox{75mm}{!}{\includegraphics{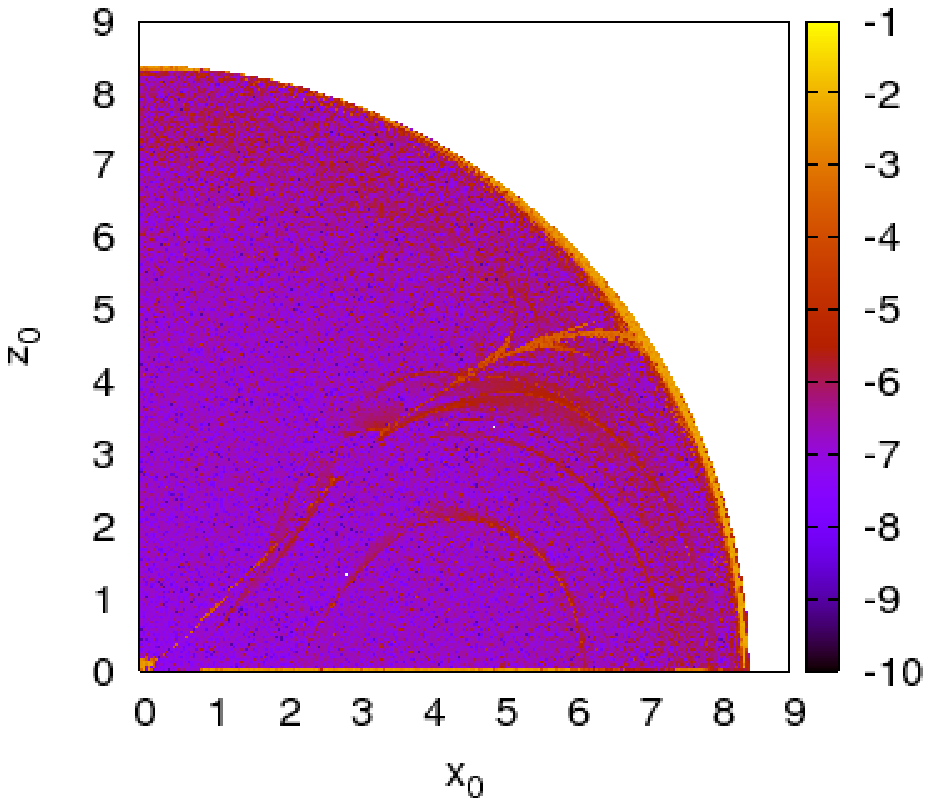}}& 
\resizebox{75mm}{!}{\includegraphics{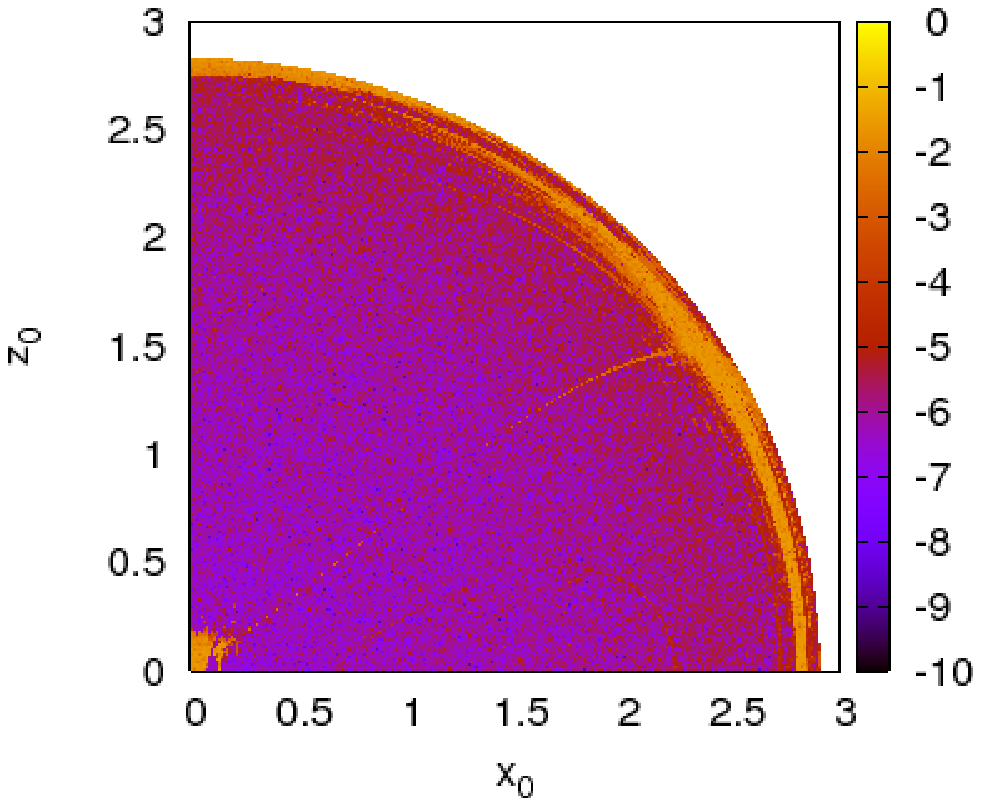}}\\
\resizebox{75mm}{!}{\includegraphics{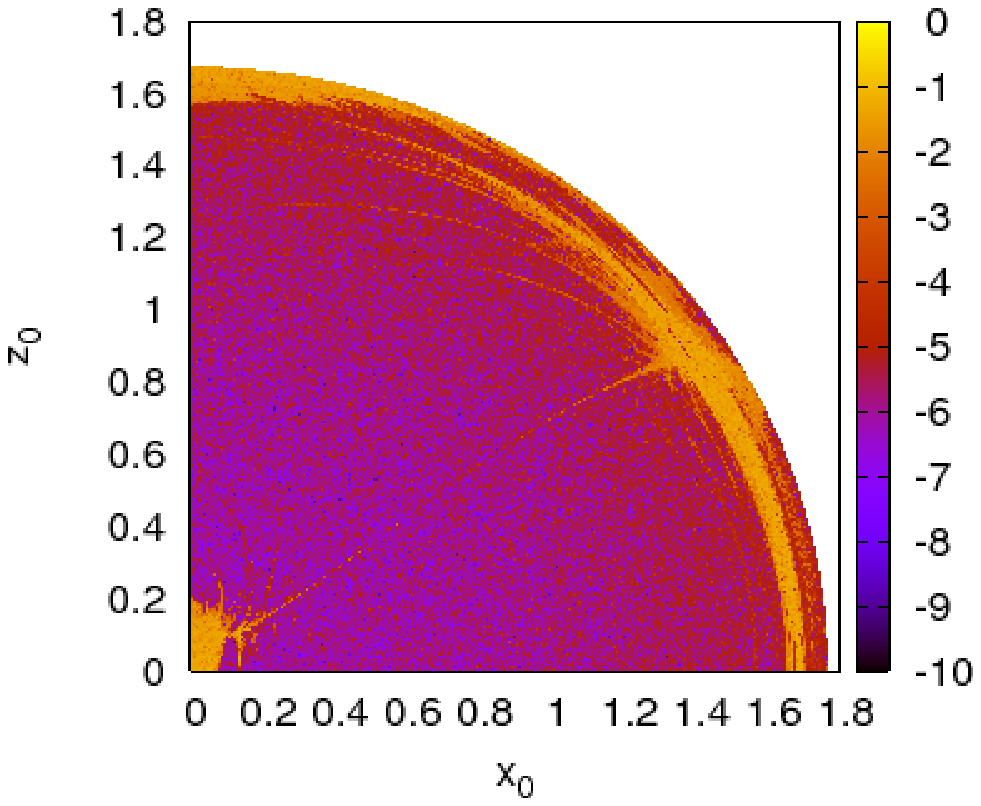}}& 
\resizebox{75mm}{!}{\includegraphics{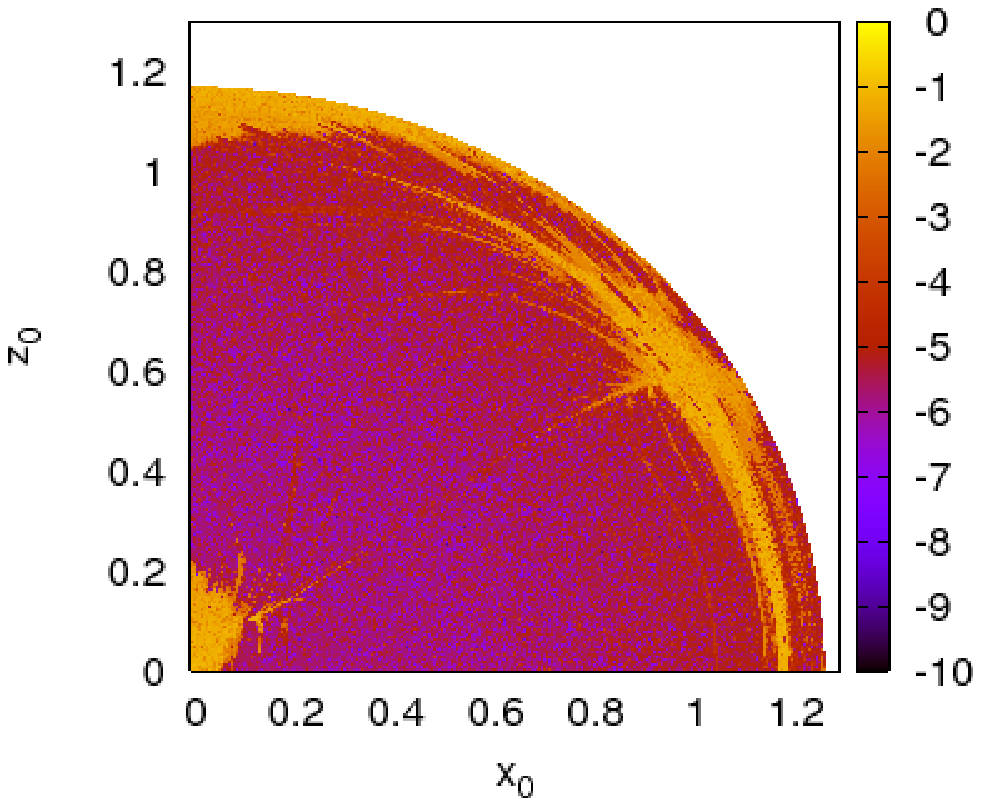}}
\end{tabular}
\caption{Phase space portraits of the SE\textit{l}LCE for the $x_0-z_0$ start space and for four energy surfaces. Top left panel: phase space portrait corresponding to the energy surface $-0.1$ for 998938 i.c. and using a total integration time of $1.17\times 10^5$ u.t. Top right panel: $-0.3$, with 998948 i.c. and $2.4\times 10^4$ u.t. Bottom left panel: $-0.5$, with 999400 i.c. and $1.2\times 10^4$ u.t. Bottom right panel: $-0.7$, with 998626 i.c. and $7\times 10^3$ u.t. Chaotic regions are identified with indicator values higher than $\sim-5$ or $\sim-4$ for the energy surfaces $-0.1$, $-0.3$ and $-0.5$ or $-0.7$, respectively. See Table \ref{tablestart-percent} for further details. Notice that the values of the indicator are in logarithmic scale.}
\label{STARTX0Z0-1}
\end{figure*}

Here, we implement the same procedure applied in Section \ref{STATIONARYS4} to study the $x_0-z_0$ start space. We obtain the phase space portraits with the SE\textit{l}LCE for four energy surfaces, calibrating the thresholds with the assistance of the LI, and we further discuss the global dynamics of the $x_0-z_0$ start space.

In order to determine the percentages of the chaotic component, we proceed like we did in Section \ref{STATIONARYS4}. We present the results ordered by energy surface on Table \ref{tablestart-percent}.

In Fig. \ref{STARTX0Z0-1}, we present the phase space portraits of the $x_0-z_0$ start space\footnote{We use a cut--off level around $10^{-10}$ to have the same scales in Fig. \ref{STARTX0Z0-1} for every energy surface. Then, the percentages of orbits that we do not consider are negligible. Those percentages are $\sim0.22$\% in the energy surface $-0.1$, $\sim0.05$\% in $-0.3$, $\sim0.02$\% in $-0.5$ and $\sim0.02$\% in $-0.7$.} for the same energy surfaces considered in Section \ref{STATIONARYS4}. Also the total integration times remain the same and the VICs used for the study guarantee reliable and stable pictures. The numbers of orbits in the samples are different and depend on the energy surface: 998938 i.c. for the energy surface $-0.1$, 998948 for $-0.3$, 999400 for $-0.5$ and 998626 for $-0.7$. 

There is a significant change in the dynamics of the $x_0-z_0$ start space with respect to the results for the $p_{x_0}-p_{z_0}$ start space. The corresponding phase space portraits presented in Fig. \ref{STARTX0Z0-1} show a dominant regular component (see Table \ref{tablestart-percent}, columns three and five) instead of the rather large chaotic component seen for the $p_{x_0}-p_{z_0}$ start space. The percentages of the smaller component have a substantial increment toward the innermost energy surface. Indeed, in this case, the expansion of the chaotic component is due to the growth of the different orbital families separatrices. This fact is clearly seen in Fig. \ref{STARTX0Z0-1}. On the top left panel of Fig. \ref{STARTX0Z0-1}, we can see a very weak separatrix that separates the short--axis tubes from the outer long--axis tubes. On the top right panel of the same figure, we can see how the former separatrix is connected to the others which separate the inner long--axis tubes and the box orbits (see \citet{PL98} for further details on the different orbital families and their locations in phase space). This becomes evident in the bottom panels of Fig. \ref{STARTX0Z0-1}. Furthermore, a strong chaotic domain appears in the crossing of the main separatrices. Therefore, as the dominant component comprises regular orbits, the main contribution to the chaotic component emerges from the separatrices of resonances.

Finally, the dynamics of the $x_0-z_0$ start space can be briefly described as a dominant regular component. The major resonances that separate the main orbital families have separatrices which grow when the negative energies increase. Furthermore, the growth of the separatrices increases the percentage of chaotic orbits very fast.

\section{Conclusions}\label{conclusions}
With this report we conclude a series of investigations (reported in \citet{MGC11}, M11 and D12) toward an efficient choice of a minimal package of techniques to study a general Hamiltonian.

Here, we apply both VICs and a SAM (the FMFT, \citet{SN97}) to study a fairly realistic dynamical model of an elliptical galaxy (the ScTS model, \citet{MCW05}).

In order to select the appropriate VICs for the experiments, we not only consider previous comparisons, but also extend them with new indicators: the SE\textit{l}LCE and the APLE \citep{LVE08}. The SE\textit{l}LCE is an efficient estimation of the lLCE by means of the MEGNO \citep{CS00} and works as its reliable alternative in case of studying big samples of orbits. The APLE has not shown advantages over the FLI/OFLI for the purposes of the experiment, but it behaves similarly. 

The SE\textit{l}LCE seems to improve the performance of the MEGNO, in particular situations. Therefore, it should be considered in the CIsF presented in D12 as a suitable alternative. 

On the other hand, we consider the performance of the FMFT as the representative SAM to compare with the VICs. The FMFT is an improvement of the FMA outlined by Laskar \citep{L90}, which is widely used by the scientific community. 

The main advantage of the SAMs is a fast computation of the frequencies. However, the performance of the FMFT as a global indicator of chaos is not as efficient as the performances shown by the VICs in the experiments. The SAMs are designed to compute the frequencies, which are quantities strictly related to regular motion. Using such techniques as global indicators of chaos involves forcing the methods to do something for which they are not designed. The VICs are based on the concept of local exponential divergence. Hence, the detection of chaos is their main purpose. A natural consequence of this is that an efficient application of those techniques is based on their complementary implementation (see Section \ref{intro} for references). The SAMs are of use to describe the resonance web while the VICs to study the interplay of regular and chaotic domains (see Section \ref{thespaces}).

Finally, the recommended CIsF for the analysis of a general Hamiltonian is composed of the MEGNO/SE\textit{l}LCE, the FLI/OFLI and the GALI$_{2N}$: the MEGNO/SE\textit{l}LCE and the FLI/OFLI as global indicators of chaos to obtain the phase space portraits and display the interaction between regular and chaotic orbits; the GALI$_{2N}$ to analyse small samples within regions of complex dynamics and regions where the chaotic component is dominant (see \citet{SBA07} and D12). Furthermore, the CIsF can be used with a SAM (e.g. the FMFT) in order to characterise the resonance web with the frequency vectors. 

\section*{Acknowledgments}
The authors want to thank Dr. J.C. Muzzio for his assistance along the investigation and the English department of FCAG--UNLP for improving the English of the reported research. Also, we are grateful to the anonymous referee because of the insightful comments made to improve the original manuscript. This work was supported with grants from the Consejo Nacional de Investigaciones Cient\'{\i}ficas y T\'ecnicas de la Rep\'ublica Argentina (CCT--La Plata) and the Universidad Nacional de La Plata.

\bsp

\label{lastpage}

\end{document}